\documentclass[aps,onecolumn,superscriptaddress,amsmath,showpacs,tightenlines]{revtex4-1}

\usepackage{graphicx}
\usepackage{subfigure}
\usepackage{natbib}
\usepackage{epsfig}
\usepackage{amsfonts}
\usepackage{amsthm,amsmath,amssymb}
\usepackage{mathrsfs}

\usepackage{epstopdf}
\usepackage{xcolor}
\usepackage[toc,page,title,titletoc,header]{appendix}

\usepackage[utf8]{inputenc}

\begin{document}
\title{Open quantum system in the indefinite environment}
\author{He Wang}
\affiliation{College of Physics, Jilin University,\\Changchun 130021, China}%
\affiliation{State Key Laboratory of Electroanalytical Chemistry, Changchun Institute of Applied Chemistry,\\Changchun 130021, China.}%
\author{Jin Wang}
\email{jin.wang.1@stonybrook.edu}
\affiliation{Department of Chemistry and of Physics and Astronomy, Stony Brook University, Stony Brook,\\NY 11794-3400, USA}%

\begin{abstract}

In this paper, we investigate the interference engineering of the open quantum system, where the environment is made indefinite either through the use of an interferometer or the introduction of auxiliary qubits. The environments are modeled by fully connected qubit baths with exact analytical dynamics. As the system passes through the interferometer or is controlled by auxiliary qubits, it is propagated along different paths or their superpositions, leading to distinct interactions with the environment in each path. This results in the superposition of the environments, which can be detected through specific measurements that retain certain coherent information about the paths. Our results demonstrate that the indefiniteness of the environment can significantly enhance the quantum correlations. However, only the statistical mixture of the influences from the environments preserves provided that the path coherence is destructed. We also examine the serviceability of the indefiniteness as a resource for teleportation and quantum parameter estimation. Additionally, we discuss how to quantify the indefiniteness and the ways in which it affects the system's dynamics from the perspective of wave-particle-entanglement-ignorance complementarity. Our study highlights the potential benefits of an indefinite environment in quantum information processing and sheds light on the fundamental principles underlying its effects.

\end{abstract}

\maketitle

\section{Introduction} \label{introduction}

The coupling of quantum systems to their surrounding environment often leads to quantum decoherence and dissipation, which have long been regarded as the most formidable obstacles to realizing quantum information processing \cite{NMA2000}. The presence of the coupling with the environment causes information to scramble into uncontrollable degrees of freedom, resulting in reduced dynamics that are often incoherent and irreversible. This loss of unitarity and associated advantages of quantum information processing, such as computational speedups, poses a significant challenge to the field. To address this issue, researchers have developed a host of techniques to tame decoherence, such as quantum error correction \cite{JR2019}, decoherence-free subspaces \cite{PZ1997,DAL1998,DAL14}, noiseless subsystems \cite{LV00,MDC06,DAL14}, feedback control \cite{HMW94,ACD00,JZ17}, dynamical decoupling \cite{LV99,DAL14}, the quantum Zeno effect \cite{WMI1990,PF02,KK05,AZC14,AZC16}, and Floquet engineering \cite{CC15,WLY19,SYB20,HW23a}, among others. Reservoir engineering has also garnered increasing attention as a viable approach to make the environment an ally instead of a foe, achieved by optimizing the noise of the environment to preserve quantum features such as coherence or entanglement \cite{XH22,HW22,KZ21,XW19,ZHW19,ZDZ15}. The initial assertion that the environment is the major enemy of all quantum technologies now needs to be re-examined. Both theoretical and experimental findings demonstrate that suitably tailored interactions with the environment can be utilized to generate or safeguard quantum resources, for instance, appropriately engineered dissipation can prepare many-body states and non-equilibrium quantum phases, and even perform quantum computation \cite{MBP99,BK08,FV09}. Therefore, reservoir engineering holds significant potential to advance the viability and scalability of quantum applications.

A novel reservoir engineering scheme, dubbed as the quantum switch \cite{GC12,II15,DE18,KG18}, has recently been proposed to mitigate the detrimental effects of the environment and preserve the quantumness of a system. By leveraging the quantum coherence of the quantum switch, it can determine the order in which a system interacts with multiple environments in an indeterminate fashion. This results in the emergence of indefinite causal order, which challenges our traditional understanding of causality in the context of quantum physics. The application of this new approach has the potential to expand our knowledge and understanding of quantum mechanics, with possible implications for the development of future quantum technologies aimed at protecting quantum systems from environmental disturbance. Studies have demonstrated that the use of the quantum switch can enhance the information capacity of communication channels \cite{DE18,NL20}, improve the fidelity of quantum teleportation \cite{CM2020,CCI2020}, and increase the quantum Fisher information in quantum metrology \cite{CM18,FCB21a,FCB21b,FCB21c,XZ22,BM2023}. Remarkably, two identical copies of a completely depolarizing channel become able to transmit information when they are combined in a quantum superposition of two alternative orders \cite{DE18}. Additionally, the experiments demonstrated that the indefinite causal order is also an valuable resource for quantum thermodynamics tasks \cite{DF20,XFN22}. It is worth noting that the indefinite causal order is achieved only if a quantum system interacts successively with two or more environments in time. However, based on space-time duality, another form of indefiniteness can be considered. This indefiniteness can be accomplished by utilizing an interferometer to generate superposed trajectories for the system. \cite{OS21,JDL22}. Alternatively, the coherent auxiliary qubits can be employed to control which environment the system occupies \cite{JDL23,MB20a,MB20b,MB21,MB22a,MB22b,CEW21}. It has been shown that this type of indefiniteness can enhance the non-Markovianity \cite{HPB18,OS21} and suppress the decoherence of an open quantum system \cite{JDL22,JDL23,MB20a,MB20b,MB21,MB22a,MB22b}.

In this paper, we investigate the concept of indefiniteness of the environment in a broader context, beyond the conventional quantum channels or simple environments composed of non-interacting harmonic oscillators studied in previous literature. Specifically, we extend our study to structured environments, such as the central spin model \cite{XZY07}, which offers the advantages of analytical solvability and experimental feasibility. Our analysis focuses on a two-qubit system interacting with an indefinite environment at first. We demonstrate that appropriate selective measurement can enhance quantum correlations due to the presence of indefiniteness. Additionally, we consider a phenomenological master equation for the path state (or the state of the control qubit) and observe that the final effect is a statistical mixture of the environmental effects, resulting from the destruction of path coherence. Furthermore, we investigate the potential applications of indefiniteness in quantum technologies, such as teleportation and parameter estimation, and show that it can serve as an advantageous resource for both tasks. We propose a measure of the indefiniteness of the environments based on the wave-particle-entanglement-ignorance complementarity \cite{WW2020}. This measure captures the local and non-local properties of the total system and can significantly alter the system's dynamics. Our study highlights the potential benefits of utilizing indefiniteness in quantum technologies.

The remainder of this paper is structured as follows. In Sec.\ref{Model and dynamics}, we provide an introduction to the concept of indefiniteness in environments and describe the model under investigation, including its analytical solution. In Sec.\ref{Results}, we present the key findings of our study, including the enhancement of quantum correlations and the advantageous use of indefiniteness in quantum technology. Finally, in Sec.\ref{Conclusion and Discussion}, we draw a conclusion based on our analysis and offer a further discussion.

\section{Model and dynamics} \label{Model and dynamics}

In this section, we begin by introducing the concept of indefinite environments, which are realized through the use of interferometers. We then present the concrete Hamiltonian for the entire system.  The dynamics of the system can be obtained exactly without any approximations. For simplicity, we adopt the natural units where $\hbar = k_B = c = 1$ throughout this paper.

\subsection{Indefinite environments for the two-qubit system as an illustration}

The system under consideration consists of two interacting qubits, which are sent through a multiport beam splitter. Subsequently, the system is prepared in the state $|\psi\rangle=|\psi_A\rangle|\psi_B\rangle=\frac{1}{\sqrt{M}}\sum_{i=1}^{M}|i_{P_A}\rangle \otimes\frac{1}{\sqrt{M}}\sum_{j=1}^{M}|j_{P_B}\rangle$, where the path Hilbert space is spanned by an orthonormal basis of states denoted by ${|i_{P_{A(B)}}\rangle }$ for the qubit $A(B)$. As the qubit $A(B)$ travels along path $i$, it locally interacts with a qubit bath $E_{i_{A(B)}}$. Importantly, when the system is prepared in a coherent superposition state, as is the case here, it interacts with all baths over the different paths in a coherent superposition manner. This opens up the possibility of engineering the environment in a controllable fashion. The total Hamiltonian governing the dynamics of the system, including the paths, the qubits, and the baths inside the interferometer, can be expressed as 

\begin{equation}\label{H_tot}
\pmb H_{tot}=\sum_{i,j=1}^{M}|i_{P_A},j_{P_B}\rangle\langle i_{P_A},j_{P_B}| \otimes \pmb H_{SE_{i,j}}  
\end{equation}

The concrete formula of the Hamiltonian governing the dynamics of the system and the baths, denoted by $H_{SE_{i,j}}$ will be presented in the next section. Upon initial inspection of Eq. \ref{H_tot}, it is apparent that it bears a striking resemblance to the discrete quantum walk \cite{VA12}. However, the dynamics of the discrete quantum walk are typically determined by an evolution operator, rather than a Hamiltonian. The control qubit or coin qubit decides the motion of the particle. When the control qubit is prepared in a coherent superposition state, the particle can traverse multiple paths simultaneously. The presence of various environments on different paths contributes to the indefiniteness of the environment. The dynamics of the system over the different paths interfere with each other. This is reflected in the evolution of the system plus the paths,

\begin{equation}\label{evol_SP}
\pmb \rho_{SP}=\frac{1}{M^2}\sum_{i,j,i',j'}|i_{P_A}\rangle\langle i_{P_A}'|\otimes|j_{P_B}\rangle\langle j_{P_B}'|\otimes \pmb \rho_{SP_{i,i',j,j'}},
\end{equation}

\begin{equation}\label{evol_S}
\pmb \rho_{SP_{i,j,i',j'}}(t)=Tr_{E_{m,n}}(e^{-i\pmb H_{SE_{i,j}}t }\pmb \rho_S(0)\otimes_{m}^{M}\otimes_{n}^{M}\pmb \rho_{E_{m,n}}(0)e^{i\pmb H_{SE_{i',j'}}t }),
\end{equation}

where $\rho_{E_{m,n}}=\rho_{E_{m}}\otimes\rho_{E_{n}}$ is the density matrix of the baths on the path $m (n)$ where the qubit $A (B)$ transverses. In Eq.\ref{evol_S}, the terms $i=i'$ and $j=j'$ describe the reduced dynamics from propagating the qubit $A$ over the path $i$, and propagating the qubit $B$ over the path $j$. The summation of these terms captures the contribution to the overall dynamics of the statistical mixture of the influence of the environment on the different paths, whereas the terms $i\neq i'$ or $j\neq j'$ capture the interference effect between the different environments over the paths.

Discarding the path information directly leads to the statistical mixture of the dynamics over the different paths. On the other hand, one can perform selective measurement on the path to leverage the interference effect. For instance, selective measurement can be chosen as

\begin{equation}
\pmb \Pi_{P,\phi}=|\psi_{A,\phi}\rangle\langle \psi_{A,\phi}|\otimes|\psi_{B,\phi}\rangle\langle \psi_{B,\phi}|,
\end{equation}

where $|\psi_{\alpha,\phi}\rangle=\frac{1}{\sqrt{M}}\sum_{i=1}^{M}e^{i\phi_i}|i_{P_\alpha}\rangle$.
$\{\phi_i\}$ determines the specific way of measurement. Physically, this process can be achieved by applying the other beam splitter with the phase shifts $\{\phi_i\}$ and selecting the output beams of the qubits \cite{OS21,JDL22}. The post-measurement state of the system becomes 

\begin{equation}\begin{split}
\tilde{\pmb\rho}_{S,\phi}&=\langle\psi_{1,\phi},\psi_{2,\phi}| \pmb \rho_{SP}|\psi_{1,\phi},\psi_{2,\phi}\rangle\\
&=\frac{1}{M^4}\sum_{i,j,i',j'}e^{i(\phi_i-\phi_{i'}+\phi_j-\phi_{j'})}\pmb \rho_{SP_{i,i',j,j'}}\\
&=\frac{1}{M^4}\pmb \sum_{i,j}\pmb\rho_{SP_{i,i,j,j}}+\frac{1}{M^4}\pmb \sum_{i\neq i'\lor j\neq j'}e^{i(\phi_i-\phi_{i'}+\phi_j-\phi_{j'})}\pmb\rho_{SP_{i,i',j,j'}},
\end{split}
\end{equation}

where the first term in the last equation describes the statistical mixture of the dynamics over the different paths and the second describes the interference effects between paths. After the normalization, the post-measurement state is obtained as $\pmb\rho_{S,\phi}=\tilde{\pmb\rho}_{S,\phi}/Tr(\tilde{\pmb\rho}_{S,\phi})$. 
It is worth noting that the indefiniteness of the environments can also be introduced by auxiliary qubits, besides the interferometry approach \cite{JDL23,MB20a,MB20b}. We will also use the language of auxiliary qubits in the following discussions.

\subsection{Central spin model and its dynamics}

In this subsection, we focus on a specific model and investigate its dynamics. The system under consideration consists of two coupled qubits, while the environments are composed of structured qubit baths with a finite number, $N$, of qubits. Spin baths are as common as harmonic oscillator baths in both real-world systems and theoretical models. In particular, spin baths are often used to simulate the effects of noise from nuclear spin, as in qubit systems based on quantum dots \cite{BU13,ZZD17}, which is considered one of the most promising approaches to quantum computing. When the qubit $A$ passes through the path $i$ and $B$ passes through the path $j$, the total Hamiltonian for the system plus the baths is

\begin{equation}
\pmb H_{SE_{i,j}}=\pmb H_{S_{i,j}}+\pmb H_{I_{i,j}}+\pmb H_{E_{i,j}}
\end{equation}

\begin{equation}
\pmb H_{S_{i,j}}=-J_{i,j}\pmb\sigma_{A,i}^z\pmb\sigma_{B,j}^z +\omega_1\pmb\sigma_{A,i}^z+\omega_2\pmb\sigma_{B,j}^z
\end{equation}

\begin{equation}
\pmb H_{I_{i,j}}=\frac{f_{i,j}}{2\sqrt{N}}\sum_{\alpha=(A,i),(B,j)}\sum_{n=1}^{N}\pmb\sigma_\alpha^x\pmb S_{n,\alpha}^x+\pmb\sigma_\alpha^y\pmb S_{n,\alpha}^y
\end{equation}

\begin{equation}
\pmb H_{E_{i,j}}=\frac{s_{i,j}}{2N}\sum_{\alpha=(A,i),(B,j)}\sum_{n\neq m}^{N}\frac{1}{2}(\pmb S_{n,\alpha}^x\pmb S_{m,\alpha}^x+\pmb S_{n,\alpha}^y\pmb S_{m,\alpha}^y)+\pmb S_{n,\alpha}^z
\end{equation}

where $\pmb H_{S_{i,j}}$, $\pmb H_{I_{i,j}}$, $\pmb H_{E_{i,j}}$ are the system Hamiltonian and the interaction Hamiltonian between the system and the bathes over the paths $i$ and $j$, and the Hamiltonian of the baths, respectively. Here, $\pmb\sigma_{\alpha}^z$ denotes the z-direction Pauli matrix of qubit $\alpha \in \{(A,i),(B,j) \}$, while $\pmb S_{m,\alpha}^z $ represents the z-direction Pauli matrix of the $m$-th qubit in bath $\alpha$. We comment on the bath Hamiltonians, which have taken homogeneous interactions for the sake of analytical clarity. Such interactions exist in practical settings like quantum spin glasses \cite{CYK14}, superconducting systems \cite{DIF08}, and others, thus this is not an oversimplified assumption. The more general system with in-homogeneous couplings can only be solved numerically. The study of fully connected qubit networks has gained increasing attention in recent years, with applications ranging from the generation of highly entangled states \cite{DIF08} to the investigation of phase transitions in the Lipkin-Meshkov-Glick model \cite{OC12,XK20,MG23},  quantum generalization of Hopfield neural networks \cite{MC06, PR18}, the study of time crystals in open quantum system \cite{AR17,GP21,HW23}, and the central spin model studied herein \cite{XZY07,XXF10,CM17,SB21,DT22,DT23}.  Furthermore, recent advances in experimental techniques have made it possible to realize these systems in the laboratory. For instance, it is possible to tailor many-body spin Hamiltonians defined over arbitrary graphs using trapped ion arrays \cite{SK12,MC21}, while a recent experiment successfully probed the dynamical phase transition in a spin model with all-to-all interactions on the superconducting platform \cite{XK20}. These advancements have opened up new avenues for investigating the dynamics of complex quantum systems and have motivated the study of fully connected qubit networks. In our subsequent analysis, we will only consider the case where $s_{ij}=s$ and $f_{ij}=f$, which means that the coupling between the system and the environments are identical and the environments Hamiltonian is uniform.

To obtain the result of Eq.\ref{evol_S}, it is necessary to consider the terms involved $i\neq i'$ or $j\neq j'$. As a result, the evolution of the qubit system plus the environments can be non-unitary. One has to compute the part as $e^{-i\pmb H_{SE_{i,j}}t }|ab\rangle|x_iy_j\rangle$, $\langle a'b'|\langle x_{i'}y_{j'}|e^{i\pmb H_{SE_{i',j'}}t }$, and then trace out the environments.  In the following, we focus on computing the term $e^{-i\pmb H_{SE_{i,j}}t }|ab\rangle|x_iy_j\rangle$. For simplicity, we omit the subscripts $i,j$, and restore them when necessary. Using the total spin operator $\pmb J_{\alpha}^k=\sum_{i=1}^{N} \pmb S_{n,\alpha}^k$ ($k=x,y,z,+,_; \alpha=A,B$), the bath Hamiltonian can be rewritten as:

\begin{equation}
\pmb H_{E}=\sum_{\alpha=A,B}s(\frac{\pmb J_{\alpha}^+\pmb J_{\alpha}^-}{N}-\frac{1}{2}).
\end{equation}

Similarly, the system--bath interaction Hamiltonian changes to

\begin{equation}
\pmb H_{I}=\frac{f}{2\sqrt{N}}\sum_{\alpha=A,B}\pmb\sigma_\alpha^x\pmb J_{\alpha}^x+\pmb\sigma_\alpha^y\pmb J_{\alpha}^y.
\end{equation}

Next, we introduce the Holstein-Primakoff transformation \cite{HP1940},
\begin{equation}\begin{split}
\pmb J_{\alpha}^+&=\sqrt{N}\pmb a_\alpha^\dagger(1-\frac{\pmb a_\alpha^\dagger \pmb a_\alpha}{N})^{1/2},\\
\pmb J_{\alpha}^-&=\sqrt{N}\pmb (1-\frac{\pmb a_\alpha^\dagger \pmb a_\alpha}{N})^{1/2}\pmb a_\alpha,
\end{split}
\end{equation}

where $[\pmb a_\alpha, \pmb a_\beta^\dagger]=\delta_{\alpha\beta}$. It is important to note that the validity of the Holstein-Primakoff transformation is limited to situations where the temperature is sufficiently low such that the system's excitations are low enough. Furthermore, the dimension of the Hilbert space of $\pmb a_\alpha^\dagger \pmb a_\alpha$ is not infinite as in the usual quantum harmonic oscillator, but rather finite and equal to $N$ due to the Holstein-Primakoff transformation. After this transformation, the Hamiltonians can be reformulated as 
\begin{equation}\label{H_E}
\pmb H_{E}=s\sum_{\alpha=A,B}(\pmb a_\alpha^\dagger \pmb a_\alpha(1-\frac{\pmb a_\alpha^\dagger \pmb a_\alpha-1}{N})-\frac{1}{2}).
\end{equation}

\begin{equation}\label{H_I}
\pmb H_{I}=f\sum_{\alpha=A,B}\pmb\sigma_\alpha^+(1-\frac{\pmb a_\alpha^\dagger \pmb a_\alpha}{N})^{1/2}\pmb a_\alpha+\pmb\sigma_\alpha^-\pmb a_\alpha^\dagger(1-\frac{\pmb a_\alpha^\dagger \pmb a_\alpha}{N})^{1/2}.
\end{equation}
The model under consideration turns into a nonlinear Jaynes-Cummings type system after applying the Holstein-Primakoff transformation. Consider the evolution of the state $|\psi(0)\rangle =|11\rangle|xy\rangle$ described by the unitary operator $\pmb U(t)=e^{-i(\pmb H_S+\pmb H_E+\pmb H_I)t}$ in the Schr\" {o}dinger picture. Assume its time evolution is $|\psi(t)\rangle =\mu_{1}(t)|11\rangle|x'y'\rangle+\mu_{2}(t)|10\rangle|x''y''\rangle+\mu_{3}(t)|01\rangle|x'''y'''\rangle$. 
We can safely omit the term describing the transition from $|11\rangle$ to $|00\rangle$ since this transition is prohibited by the Hamiltonians. Introduce three bath operators $\pmb A(t),\pmb B(t),\pmb C(t) $ such that $\pmb A(t)|xy\rangle=\mu_{1}(t)|x'y'\rangle, \pmb B(t)|xy\rangle=\mu_{2}(t)|x''y''\rangle, \pmb C(t)|xy\rangle=\mu_{3}(t)|x'''y'''\rangle$. Therefore, we obtain 
\begin{equation}
|\psi(t)\rangle=\pmb A(t)|11\rangle|xy\rangle+\pmb B(t)|10\rangle|xy\rangle+\pmb C(t)|01\rangle|xy\rangle.
\end{equation}

By applying the  Schr\" {o}dinger equation corresponding to the total evolution $\frac{d}{dt}|\psi(t)\rangle =-i(\pmb H_{S}+\pmb H_{I}+\pmb H_{E})|\psi(t)\rangle$, we obtain the expression

\begin{equation}\begin{split}\label{eq16}
\frac{d\pmb A(t)}{dt}&=-i(-J+\omega_1+\omega_2+\pmb H_{E})\pmb A(t)-if(1-\frac{\pmb a_2^\dagger \pmb a_2}{N})^{1/2}\pmb a_2\pmb B(t)-if(1-\frac{\pmb a_1^\dagger \pmb a_1}{N})^{1/2}\pmb a_1 \pmb C(t)\\
\frac{d\pmb B(t)}{dt}&= -i(J+\omega_1-\omega_2+\pmb H_{E})\pmb B(t)-if\pmb a_2^\dagger(1-\frac{\pmb a_2^\dagger \pmb a_2}{N})^{1/2}\pmb A(t))\\
\frac{d\pmb C(t)}{dt}&= -i(J-\omega_1+\omega_2+\pmb H_{E})\pmb C(t)-if\pmb a_1^\dagger(1-\frac{\pmb a_1^\dagger \pmb a_1}{N})^{1/2}\pmb A(t))
\end{split}
\end{equation}

The Hamiltonian given by Eq. \ref{H_E} and \ref{H_I} is of Jaynes-Cumming type. One can obtain a block-diagonalized form of this Hamiltonian by considering the dressed state subspace, which is defined as the space spanned by the states $|a,b,x,y\rangle$, where $a$, $b$, $x$, and $y$ are non-negative integers satisfying the constraint $a+b+x+y=const$. In this dressed state basis, the above system of ordinary differential equations can be written using c-numbers, which greatly simplifies the problem \cite{XZY07,XXF10,CM17,DT22}. To this end, we can introduce an operator transformation given by $\pmb A(t)=\pmb A_1(t)$, $\pmb B(t)=\pmb a_2^{\dagger}\pmb B_1(t)$, and $\pmb C(t)=\pmb a_1^{\dagger}\pmb C_1(t)$. With this transformation, the equation of motion given by Eq. \ref{eq16} can be expressed as

\begin{equation}\begin{split}
\frac{d\pmb A_1(t)}{dt}&=-i(-J+\omega_1+\omega_2+s(\pmb n_1(1-\frac{\pmb n_1-1}{N})-\frac{1}{2})+s(\pmb n_2(1-\frac{\pmb n_2-1}{N})-\frac{1}{2}))\pmb A_1(t)\\
&-i f(1-\frac{\pmb n_2} {N})^{1/2}(1+\pmb n_2)\pmb B_1(t)-if(1-\frac{\pmb n_1}{N})^{1/2}(1+\pmb n_1)\pmb C_1(t),\\
\frac{d \pmb B_1(t)}{dt}&= -i(J+\omega_1-\omega_2+s(\pmb n_1(1-\frac{\pmb n_1-1}{N})-\frac{1}{2}))\pmb B_1(t)+ s((\pmb n_2 +1)(1- \frac{\pmb n_2 }{N})-\frac{1}{2}) \pmb B_1(t)-if(1-\frac{\pmb n_2}{N})^{1/2}\pmb A_1(t)),\\
\frac{d\pmb C_1(t)}{dt}&= -i(J-\omega_1+\omega_2+s((\pmb n_1 +1)(1- \frac{\pmb n_1 }{N})-\frac{1}{2}))\pmb C_1(t)+s(\pmb n_2(1-\frac{\pmb n_2-1}{N})-\frac{1}{2}))\pmb C_1(t)-if(1-\frac{\pmb n_1}{N})^{1/2}\pmb A_1(t)).
\end{split}
\end{equation}

where $\pmb n_\alpha=\pmb a_\alpha^\dagger\pmb a_\alpha$ are number operators corresponding to bosonic operators of the two baths respectively. From the above equation, we observe that $\pmb A_1(t), \pmb B_1(t), \pmb C_1(t)$ are the only operators involved in the number operators of the two baths. Work in the Fock space of two baths, all operators can be replaced with c-numbers. We obtain the dynamics in the subspace by solving the first-order ordinary differential equations.

The similar procedures can be done for $|\chi(t)\rangle=\pmb U(t)|00\rangle=\pmb D(t)|00\rangle|xy\rangle +\pmb E(t)|01\rangle|xy\rangle+\pmb F(t)|10\rangle|xy\rangle, |\xi(t)\rangle=\pmb U(t)|01\rangle=\pmb G(t)|01\rangle|xy\rangle +\pmb H(t)|00\rangle|xy\rangle+\pmb I(t)|11\rangle|xy\rangle$ and $|\varrho(t)\rangle=\pmb U(t)|10\rangle=\pmb J(t)|10\rangle|xy\rangle +\pmb K(t)|11\rangle|xy\rangle+\pmb L(t)|00\rangle|xy\rangle$ by introducing 
$\pmb D(t)=\pmb D_1(t), \pmb E(t)=\pmb a_2\pmb E_1(t), \pmb F(t)=\pmb a_1\pmb F_1(t); \pmb G(t)=\pmb G_1(t), \pmb H(t)=\pmb a_2^\dagger\pmb H_1(t), \pmb I(t)=\pmb a_1\pmb I_1(t)$, and $\pmb J(t)=\pmb J_1(t), \pmb K(t)=\pmb a_2\pmb K_1(t), \pmb L(t)=\pmb a_1^\dagger\pmb L_1(t)$ respectively. After these procedures, we can calculate the Eq.\ref{evol_SP} term by term through tracing out the baths. The results are shown in Appendix.\ref{appendix_A}.

\section{Results}\label{Results}

Once we have the constructed density matrix, we will begin by examining the dynamics of quantum correlations. We will introduce some key measures of quantum correlations. Coherence, which is essential to interference phenomena, plays a fundamental role in quantum physics by enabling applications that are impossible within classical mechanics. Coherence can be measured as~\cite{Quantifying_Coherence}
\begin{equation}
\label{eq:12}
\mathscr{C}_{l_{1}}=\sum_{i\neq j}\mid\rho_{ij}\mid.
\end{equation}

Quantum entanglement is a crucial resource for achieving various quantum information processing tasks like teleportation \cite{Teleporting}, quantum key distribution \cite{EA1991}, quantum computing \cite{TS04,MAN00}, and more. Among the numerous measures of entanglement for a two-qubit system, concurrence is widely utilized in various contexts. The concurrence of a two-qubit mixed state $\rho$ is defined as \cite{HR09,SH97}

\begin{equation}
\label{eq:13}
\mathscr{C}=Max(0,\lambda_{1}-\lambda_{2}-\lambda_{3}-\lambda_{4}).
\end{equation}

Here, $\lambda_{i}$ denotes the square root of the $ith$ eigenvalue, in descending order of the matrix $\rho\widetilde{\rho}$ where $\widetilde{\rho}=(\sigma_{2}\bigotimes\sigma_{2})\rho^{T}(\sigma_{2}\bigotimes\sigma_{2})$, and $T$ represents transposition.

Another significant quantum correlation measure is the quantum discord~\cite{HO01}. It assesses the non-classical correlation between the two subsystems of a quantum system. The discord encompasses correlations stemming from quantum physical effects, but it does not necessarily involve the concept of quantum entanglement. In reality, it represents a distinct form of quantum correlation compared to entanglement, as separable mixed states (i.e., those without entanglement) can exhibit non-zero quantum discord. At times, it is also recognized as a measure of the quantumness of the correlation functions. The geometric discord of a bipartite quantum state is defined as~\cite{GD}
\begin{equation}
\label{eq:14}
\mathscr{D}(\rho)=\min{||\rho-\rho_{0}||^2}_{\rho_{0}\in\Omega}
\end{equation}
where $\Omega$ represents the set of zero-discord states and $||X-Y||^2=Tr(X-Y)^{2}$ is the square norm in the Hilbert-Schmidt space.  This can be calculated for any two-qubit state. The density matrix for any two-qubit state can be expressed as follows::

\begin{equation}
\label{eq:15}
\rho_{AB}=\frac{1}{4}(I_{a}\bigotimes I_{b}+\sum_{i=1}^{3}(a_{i}\sigma_{i}\bigotimes I_{b}+I_{a}\bigotimes b_{i}\sigma_{i})+\sum_{i,j=1}^{3}C_{ij}\sigma_{i}\bigotimes\sigma_{j}).
\end{equation}
The geometric discord is given as~\cite{GD}
\begin{equation}
\label{eq:16}
\mathscr{D}(\rho)=\frac{1}{4}(||a||^2+||C||^2)-\lambda_{max})
\end{equation}
where $\lambda_{max}$ is the maximum eigen-value of the $aa^{T}+CC^{T}$. $a$ is the vector composed of $a_{i}$, and $C$ is the matrix composed of $C_{ij}$.

Furthermore, apart from these quantum correlation measures, it is also possible to quantify them in relation to a specific use, such as the fidelity of teleportation. Quantum teleportation utilizes quantum entanglement as a teleportation channel to transfer quantum states from sender to receiver without sending qubits \cite{Teleporting}. The optimal fidelity of a general mixed state $\rho$ for all strategies is given by \cite{horodecki1996teleportation}

\begin{equation}
\label{eq:17}
F=\frac{1}{2}(1+\frac{1}{3}Tr(\sqrt{C^{\dagger}C})).
\end{equation}
where $C$ has been mentioned in Eq.\ref{eq:15}. It is important to note that the state that forms the quantum channel is advantageous for teleportation only when the fidelity exceeds the threshold of 2/3, which represents the upper limit for classical teleportation \cite{popescu1994bell}.

\subsection{Boosting the quantum correlations with the indefiniteness}

In this subsection, we consider a scenario in which all baths along the paths are identical, and the inter-qubit coupling $J_{ij}$ is a constant. We will generalize to more general cases in the next subsection. Furthermore, we restrict the shifting phase to take only the values $\phi_i\in\{0,\pi\}$ in this paper. In this context, statistical mixtures are not effective, and the new dynamics are entirely determined by the interference effect. Specifically, we consider the selective measurement $\pmb \Pi_{P,\phi_n}=|\psi_{A,\phi_n}\rangle\langle \psi_{A,\phi_n}|\otimes|\psi_{B,\phi_n}\rangle\langle \psi_{B,\phi_n}|$, where $n$ denotes the number of phase shifts that take the value $\pi$. The unnormalized post-measurement state is then given by

\begin{equation}
\tilde{\pmb \rho}_{S,\phi_n}(t)=\frac{1}{M^2} \pmb \rho_{SP_{i,j,i,j}}(t)+\frac{2((M-2n)^2-M)}{M^3}\pmb \rho_{SP_{i,j,i,j'}}(t)+\frac{(M^2 + 4 n^2 - M (1 + 4 n))^2}{M^4}\pmb \rho_{SP_{i,j,i',j'}}(t),
\end{equation}

where $i,j,i',j'\in\{0,M-1\}$, and $i\neq i', j\neq j'$. One can find that $\pmb \rho_{S,\phi_{\frac{M}{2}-n}}(t)=\pmb \rho_{S,\phi_{\frac{M}{2}+n}}(t)$ for $n\in\{0,\frac{M}{2}-1\}$. Note that at $t=0$, we have

\begin{equation}
\pmb \rho_{S,\phi_n}(0)=\left\{
\begin{aligned}
&\pmb \rho_{S}(0) \quad n\neq \frac{M}{2}\\
&0 \quad \quad \quad n= \frac{M}{2}\\
\end{aligned}\right.
\end{equation}

\begin{figure}[!ht]
    \centering
\includegraphics[width=7 in]{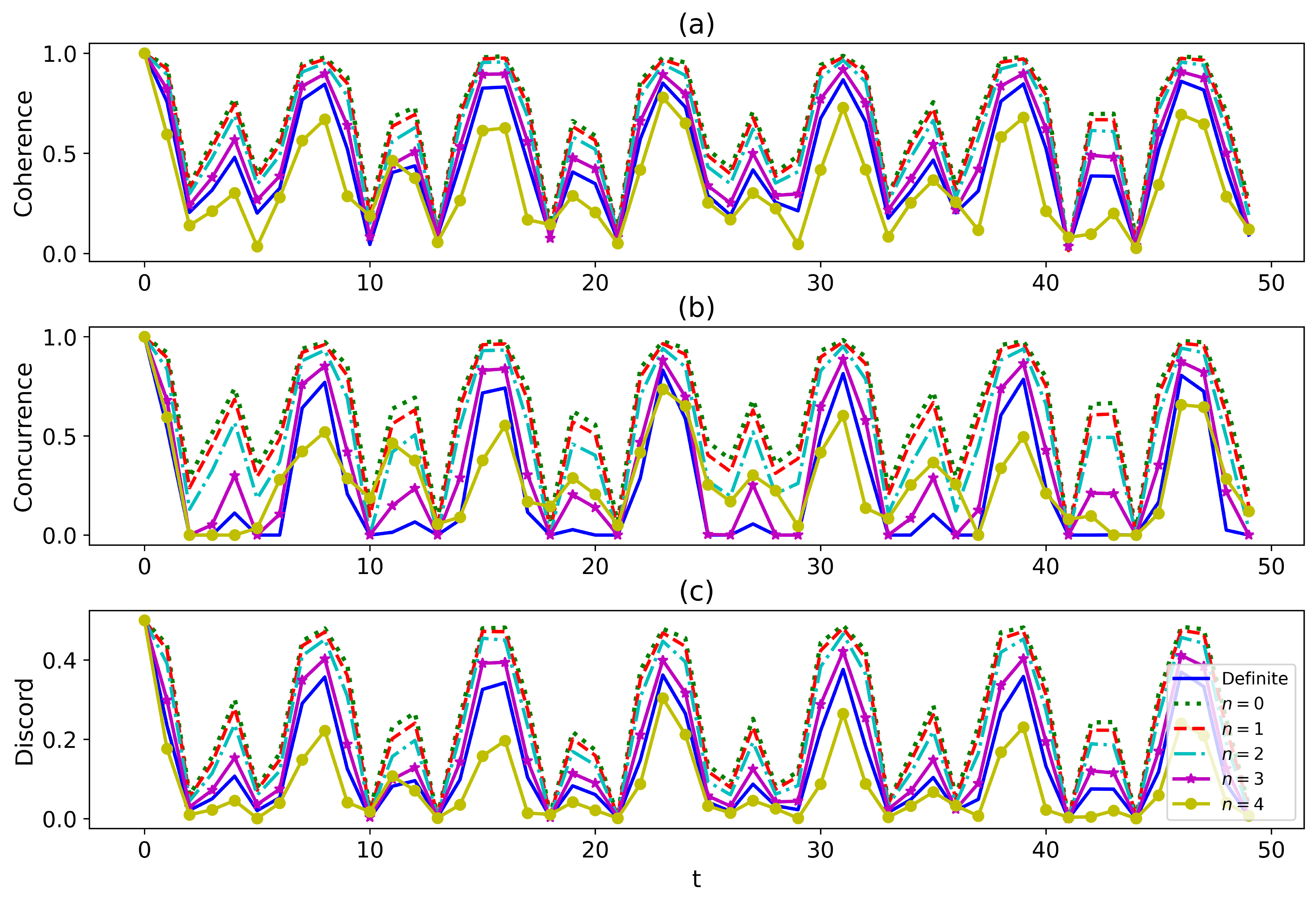}
\caption{\label{fig1} The snap plot of the dynamics of (a) coherence, (b) entanglement, (c) quantum discord with various numbers of the phase shifts. The parameters are $\omega_1=1.2, \omega_2=0.8, J=0.5, h=1, s=0.5$. The temperature is $T=0.3$. There are $M=10$ paths. The single bath is composed of $N=100$ interacting qubits.}
\end{figure}

\begin{figure}[!ht]
    \centering
\includegraphics[width=7 in]{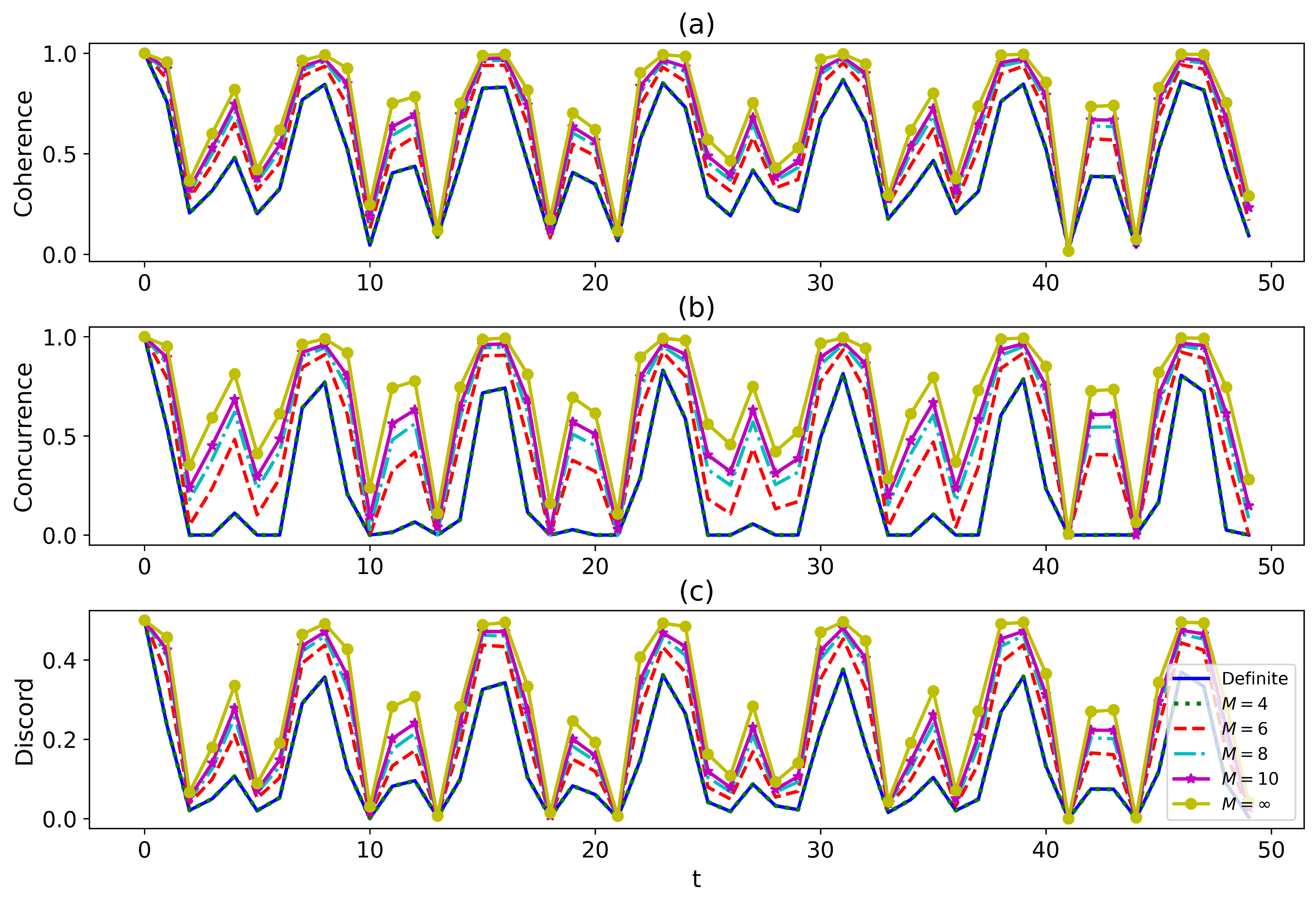}
\caption{\label{fig2} The snap plot of the dynamics of (a) coherence, (b) entanglement, (c) quantum discord with various numbers of the paths. The parameters in our analysis are $\omega_1=1.2, \omega_2=0.8, J=0.5, h=1, s=0.5$. The temperature is set to $T=0.3$. There are $n=1$ phase shifts. The single bath is composed of $N=100$ interacting qubits.}
\end{figure}

That implies the physical state is totally erased by destructive interference. We will exclude this case in this paper. With the exact reduced dynamics of the central spin obtained, in the following, we study the temporal variation of non-classical properties, viz. quantum coherence, entanglement, and quantum discord of the system. The system is prepared in the Bell state $|\Phi^+\rangle=\frac{1}{\sqrt{2}}(|11\rangle+|00\rangle)$ and then evolve. In Fig. \ref{fig1}(a,b,c), we compute the coherence, entanglement, and quantum discord, respectively, and examine the influence of the selective measurements on the final results. The coherence, entanglement, and quantum discord exhibit similar profiles, as depicted by the solid blue line.  The non-classical properties considered in our analysis exhibit long-time non-monotone oscillations. Previous studies suggest that these non-monotone oscillations are the characteristics of the non-Markovianess \cite{XZY07,CM17,DT22}. Now we investigate the dynamics when the environment is indefinite. The dynamics are only dependent on the number of phase shifts, rather than the position, due to the uniformity of the environment. We examine the behavior of the system under selective measurements and observe that, for most measurements, quantum correlations are enhanced when the environment is indefinite in comparison with a definite environment. Our findings are consistent with the previous studies \cite{MB21,MB22b}, which suggest that quantum correlations can be enhanced by environmental indefiniteness.  However, we also observe instances where the quantum correlations are lower in an indefinite environment than that in a definite one, as shown by the yellow solid line with star markers in Fig. \ref{fig1}(a,b,c). This indicates the presence of both constructive and destructive interference, just as two monochromatic beams interfere. Even so, our results suggest an effective approach that boosts quantum correlations through interferometry engineering. In a recent study \cite{JDL22}, it was shown that when an infinite number of paths is added to a system, the system's evolution can become frozen for certain states, resulting in the quantum Zeno effect \cite{WMI1990}. Unlike the usual quantum Zeno effect, which requires a temporal sequence of measurements on the system, the quantum Zeno effect described in \cite{JDL22} - termed the space-time dual quantum Zeno effect - only requires a single measurement on the $N$ paths taken by the system in an interferometer. We also examine the dynamics of the quantum correlations when the system has various numbers of superposition paths in Fig.\ref{fig2}(a,b,c). There is no frozen evolution of the system even as the number of paths reaches infinity in our model as shown in \ref{fig2}(a,b,c). However, the coherence, entanglement, and quantum discord are significantly enhanced by increasing the paths for the fixed finite phase shifts. Particularly, when the number of paths tends to be infinite, the quantum correlations achieve the maximum values at certain moments.

\subsection{Indefinite system in the indefinite environment}

In this subsection, we investigate a general scenario where both the system and the environment are indefinite. To this end, we consider the interaction between two system qubits to be position-dependent, with $J_{i,j}=\frac{1}{\gamma(i-j)^2 + d}$. This implies that the strength of the interaction between the qubits is not uniform and varies with their position. Additionally, we assume that the temperatures of the baths on the different paths are distinct. Consequently, when the system passes through the interferometer, we do not know the strength of the interaction between the qubits or the temperature of the bath to which the qubit is coupled. We restrict our analysis to the case of three-slit interference, and it is easy to generalize to other more general cases.

\begin{figure}[!ht]
    \centering
\includegraphics[width=7 in]{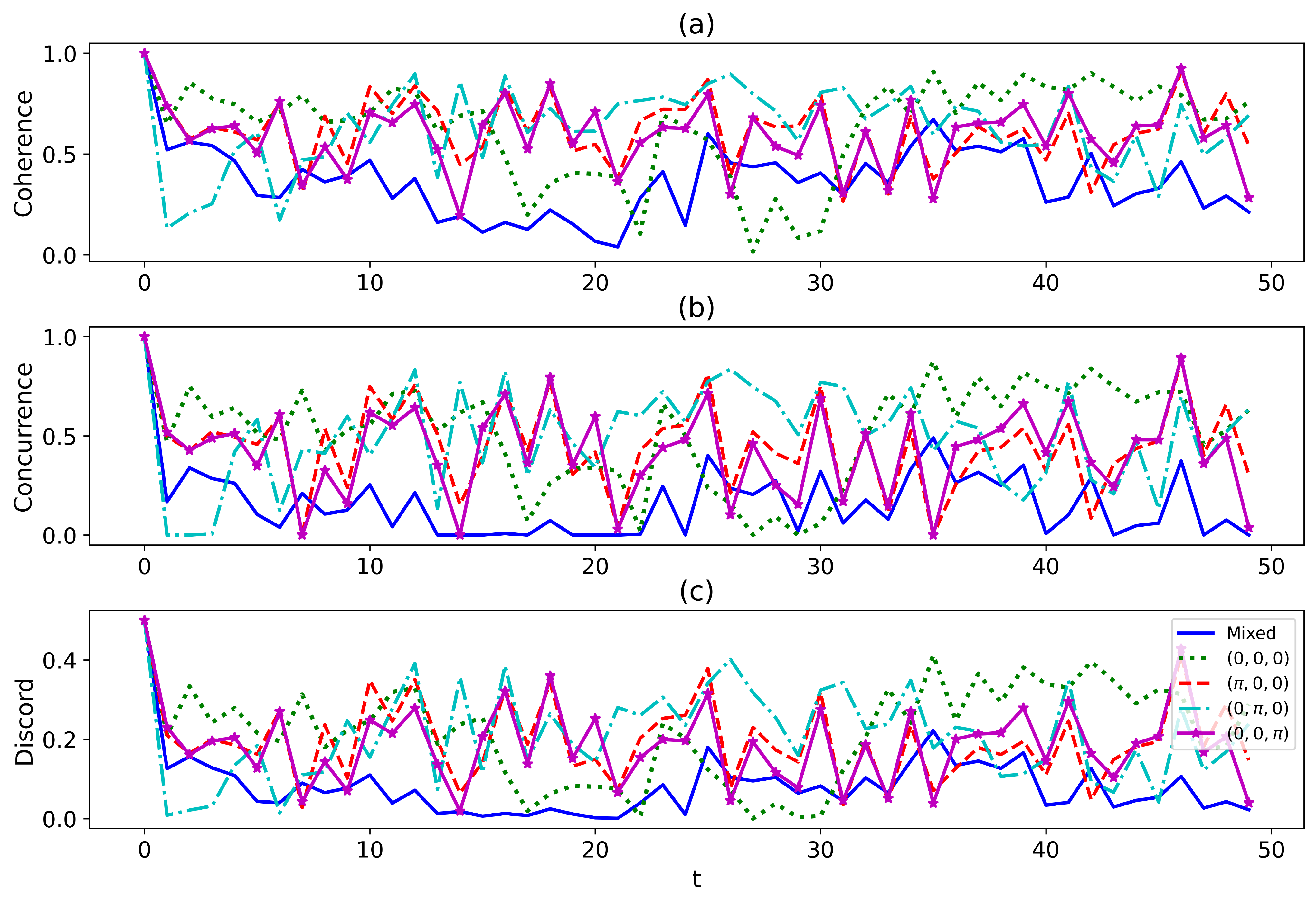}
\caption{\label{fig3} The snap plot of the dynamics of (a) coherence, (b) entanglement, (c) quantum discord. The parameters in our analysis are $\omega_1=2, \omega_2=1.5, \gamma=0.1, d=0.5, h=1.2, s=0.8$. The temperatures on the different paths are set to $T_0=0.1, T_1=0.3, T_2=0.5$. The single bath is composed of $N=100$ interacting qubits. The legend on the figure describes the phase shifts on the different paths. }
\end{figure}

In Fig.\ref{fig3}, we examine the dynamics of the quantum correlations. The profile of the coherence, entanglement, and quantum discord is similar. The non-classical properties exhibit long-time non-monotone variations. The blue solid line depicts the result obtained by tracing out the path information directly at the end, leading to a statistical mixture of the system on different paths.  In contrast, the other lines represent scenarios where selective measurements are performed at the end, preserving the path interference and maintaining the indefiniteness of the system and environment. From the observations on Fig.\ref{fig3}(a,b,c), although the mixed case occasionally exhibits larger quantum correlations than the indefinite case, the indefinite case generally dominates. Therefore, the indefiniteness can enhance the quantum correlations in general cases.

\begin{figure}[!ht]
    \centering
\includegraphics[width=7 in]{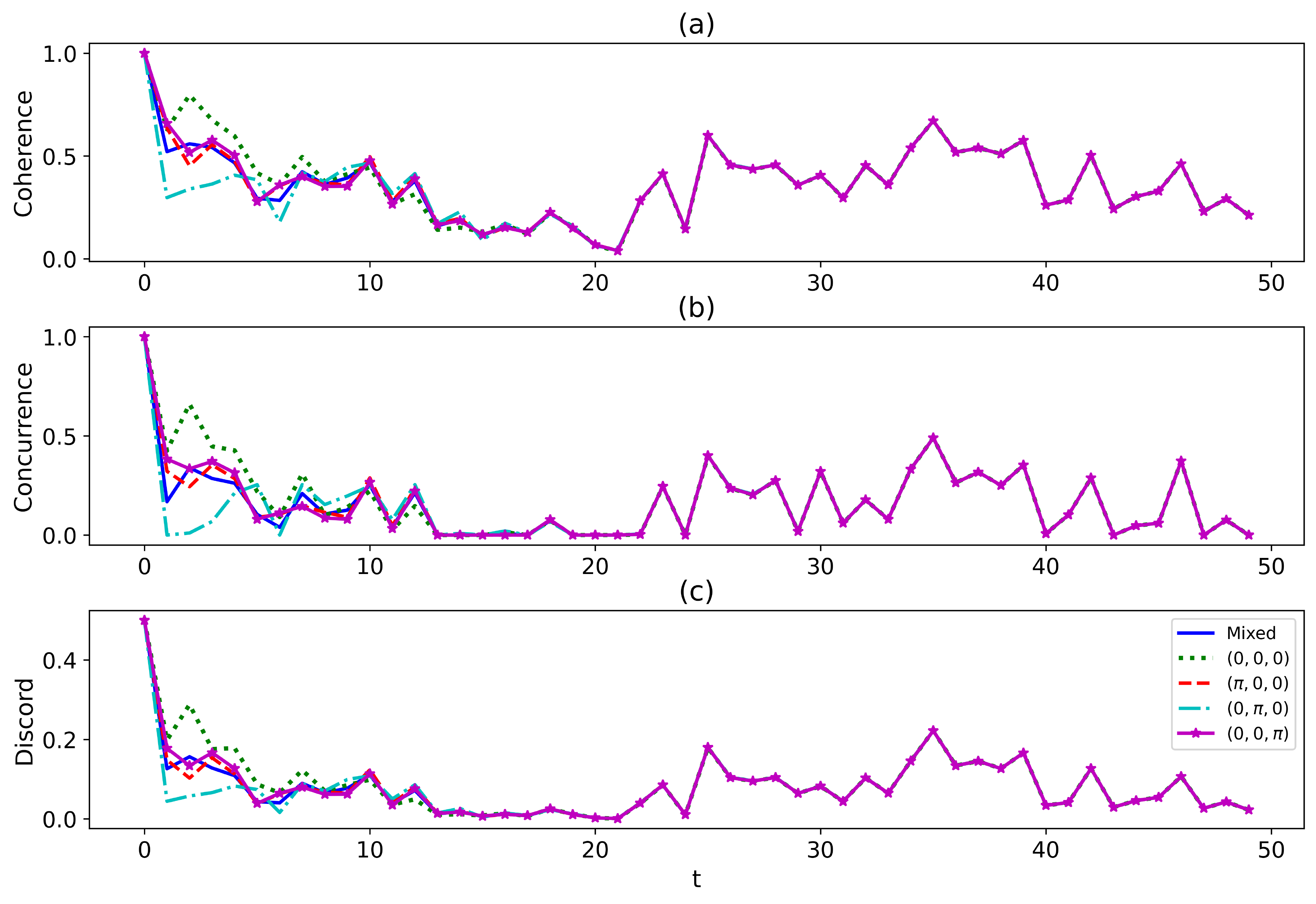}
\caption{\label{fig4} The snap plot of the dynamics of (a) coherence, (b) entanglement, (c) quantum discord when the path state suffers from decoherence. The parameters are similar to Fig.\ref{fig3}. The dissipative rate $\Gamma=0.5$. }
\end{figure}

In addition to the effects of environmental and system indefiniteness, one may also wonder what happens if there is decoherence in the path space. In this regard, we consider the following phenomenological master equation,

\begin{equation}\label{master_eq}
\frac{d \pmb\rho_{p_\alpha}(t)}{dt}=-i[\pmb H_{p_\alpha},\pmb \rho_{p_\alpha}]+\Gamma\sum_{i=0}^2 \pmb L_{i\alpha} \pmb\rho_{p_\alpha}\pmb L_{i\alpha}^\dagger-\frac{1}{2}\{\pmb L_{i\alpha}^\dagger\pmb L_{i\alpha},\pmb\rho_{p_\alpha}\},
\end{equation}

where the Lindblad operators are $\pmb L_{0\alpha}=|0\rangle_\alpha\langle1|,\pmb L_{1\alpha}=|1\rangle_\alpha\langle2|, \pmb L_{2\alpha}=|2\rangle_\alpha\langle0|$ and $\pmb H_{p_\alpha}=0$.  In Fig.\ref{fig4}, we plot the dynamics of the (a) coherence, (b) entanglement, and (c) quantum discord when the path state is governed by Eq.\ref{master_eq}. The dynamics of coherence, entanglement, and quantum discord have similar behaviors, with all lines converging to the blue solid line promptly. This indicates that the path interference is lost due to the decoherence, and only the result of the statistical mixture emerges at the end.  It has been shown that even the statistical mixture of the two environments can yield exotic phenomena \cite{HPB18}. However, it is not the case for the environments being uniform.

\subsection{Teleportation in the indefinite environment}

\begin{figure}[!ht]
    \centering
\includegraphics[width=7 in]{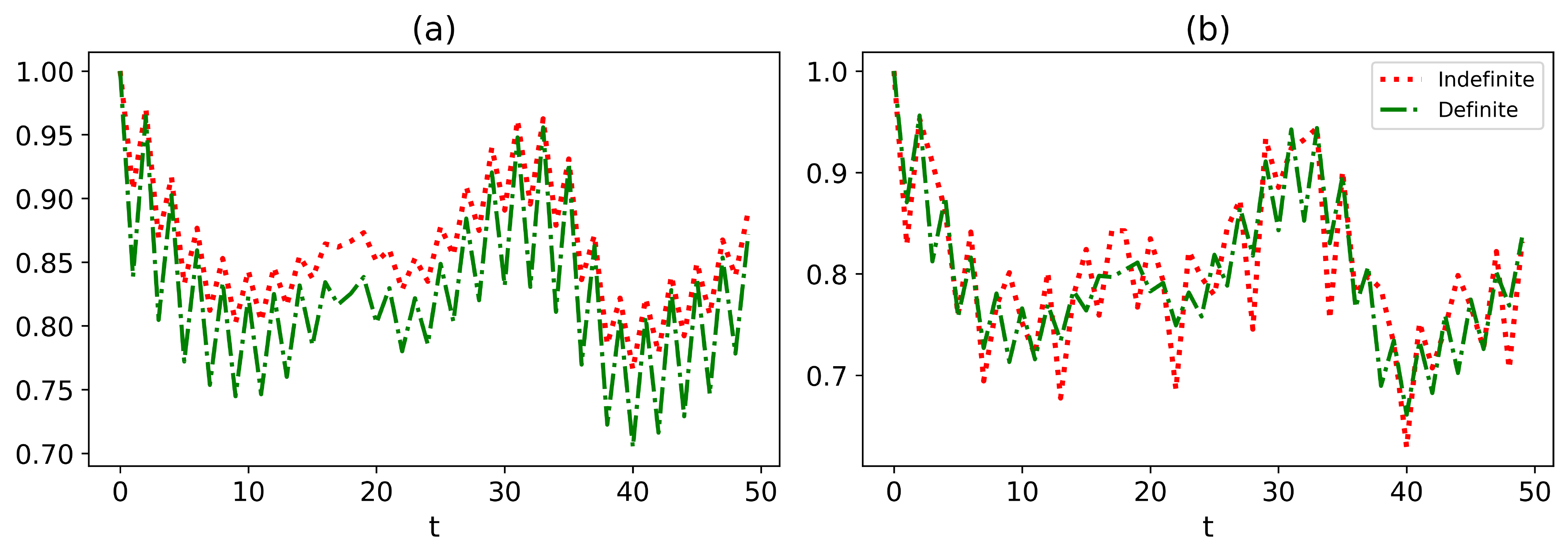}
\caption{\label{fig5} The maximum teleportation fidelity of two protocols versus time, (a) standard protocol and (b) new protocol. The parameters in our analysis are $\omega_1=2, \omega_2=1.5, \gamma=0.1, h=1.2, s=0.8$. The temperatures on the different paths are set to $T_0=0.1, T_1=0.8$. The single bath is composed of $N=100$ interacting qubits. }
\end{figure}

We study teleportation in the indefinite environment in this subsection. We consider a scenario where two qubits, denoted as qubit A and qubit B, are prepared in the maximally entangled Bell state $|\Phi^+\rangle=\frac{1}{\sqrt{2}}(|11\rangle+|00\rangle)$ and sent to the remote locations through the environments $E_0$ and $E_1$, respectively, depending on the state of the control qubit C. Specifically, if C is in the state $|0\rangle$, qubit A is sent through environment $E_0$ and qubit B through environment $E_1$. Conversely, if C is in the state $|1\rangle$, qubit A is sent through environment $E_1$ and qubit B through environment $E_0$. If the control qubit C is in a coherent state $c_1|1\rangle+c_0|0\rangle$, the environments through which qubits A and B travel are indefinite, and the qubits do not couple anymore. The baths on the two paths are both composed of interacting qubits as described Eqn.\ref{H_E}. After a lengthy but straightforward calculation, the evolution of the density matrix of the three qubits (ABC) is presented in Appendix.\ref{appendix_B}. We make the selective measurements on the control qubit, which are described by $|\pm\rangle\langle\pm|  
  (|\pm\rangle=\frac{1}{\sqrt{2}}(|0\rangle\pm|1\rangle))$ to extract certain information from the control qubit. The post-measurement state is

\begin{equation}
\pmb \rho_{AB}(t,\pm)=\frac{1}{4P(t,\pm)}(\pmb\rho_{ABC}^{0,0}+\pmb\rho_{ABC}^{1,1}\pm\pmb\rho_{ABC}^{0,1}\pm\pmb\rho_{ABC}^{1,0}),
\end{equation}

with probability $P(t,\pm)=\frac{1}{2}(1\pm\frac{1}{2}Tr_{A,B}[\pmb\rho_{ABC}^{0,1}+\pmb\rho_{ABC}^{1,0}])$. Using the  quantum state, we perform the standard teleportation protocol to teleport an unknown state.  The maximum teleportation fidelity
of a two-qubit state over all strategies given by Eqn.\ref{eq:17} is plotted in Fig.\ref{fig5}(a). The fidelity oscillates non-monotonically for both definite and indefinite cases. However, the indefinite case dominates at all moments. It has also been shown that the indefinite causal order (or quantum switch) can improve the fidelity of quantum teleportation \cite{CM2020,CCI2020}. However, they only consider simple depolarizing channels. Here we uncover that the indefiniteness of the complex environment can also improve the fidelity remarkably.

In addition to the standard teleportation protocol, we also investigate a new teleportation protocol where the teleported qubit participates in the control of which paths A and B pass through. Without loss of generality, we initialize the control qubit at $\frac{1}{\sqrt{2}}(|1\rangle+|0\rangle)$. We then perform the projective measurements $\{|\Phi^\pm\rangle\langle\Phi^\pm|,|\Psi^\pm\rangle\langle\Psi^\pm| \}$ (where $|\Phi^\pm\rangle=\frac{1}{\sqrt{2}}(|11\rangle\pm|00\rangle,|\Psi^\pm\rangle=\frac{1}{\sqrt{2}}(|01\rangle\pm|10\rangle)$) on the party AC. To achieve maximum fidelity, we apply unitary transformations on the different post-measurement states, as presented in Appendix C. In Fig. \ref{fig5}(b), we plot the fidelity of two protocols - the standard protocol (green dashed-dotted line) and the new protocol (red dotted line). Our analysis reveals that the two protocols have similar fidelity, but there are larger fluctuations in performance for the indefinite case. By examining the time-averaged fidelity, we observe that the indefinite case is enhanced by approximately 1\% compared to the definite case. We expect the new protocol may work better in some other types of environments.

\subsection{Enhancing the quantum parameter estimation with the indefiniteness}

We examine quantum parameter estimation in the indefinite environment in this subsection. In parameter estimation theory, Fisher information is crucial for establishing an upper bound on estimate accuracy \cite{2,fish}. Let $x$ be a random variable with probability distribution $p_{\theta}(x)$, where $\theta$ is the distribution parameter. With a set of observed random variables, we can estimate the parameter $\theta$ based on the observed distribution. However, this estimation may deviate from the true value, and the variance of $\theta$ is determined by the Cram\'er-Rao (lower) bound.

\begin{equation}\label{cfi}
  \text{Var}(\theta_{\rm est})\ge \frac{1}{J_\theta^{(M)}}\,,
\end{equation}
where $ \text{Var}(\theta_\text{est}):=\langle(\theta_\text{est}-\theta)^2\rangle$ is the variance of $\theta_\text{est}$ and the Fisher information $J_\theta$ is defined as
\begin{equation}
  J_\theta^{(M)}= \int d^M x \frac{1}{p_\theta(x)}\left(\frac{\partial p_\theta(x)}{\partial \theta}\right)^2\,. 
\end{equation}

In the context of quantum systems, the state of the system is described by a density matrix $\rho_{\theta}$ rather than a probability distribution function. After performing measurements on the system, random data are collected and can be used to infer the parameter $\theta$ of the system. Information about the system is obtained using a positive-operator-valued measurement (POVM), represented by a positive operator $\pmb M_x$ corresponding to the possible measurement outcomes labeled by $x$. The classical probability $p_{\theta}(x)$ in Eqn. \ref{cfi} is defined as $p_{\theta}(x)=Tr(\pmb\rho_{\theta}(x)\pmb M_x)$. By introducing the symmetric logarithmic derivative $L_{\rho_\theta}$,

\begin{equation}
  \frac{\partial\pmb\rho_\theta}{\partial \theta}=\frac{1}{2}(\pmb\rho_\theta \pmb L_{\rho_\theta}+\pmb L_{\rho_\theta}\pmb\rho_\theta)\,,
\end{equation}
where
$\pmb L_{\rho_\theta}=2\int_{0}^{\infty} ds e^{-s \pmb\rho_\theta} \frac{\partial\pmb\rho_\theta}{\partial \theta}e^{-s \pmb\rho_\theta}\,.$
The Fisher information for $M=1$ in Eqn.\ref{cfi} can be written as
\begin{equation}
  J_\theta=\int dx\frac{1}{Tr(\pmb\rho_\theta \pmb M_x)}\left(Tr\left(\frac{1}{2}(\pmb\rho_\theta \pmb L_{\pmb\rho_\theta}+\pmb L_{\pmb\rho_\theta}\pmb\rho_\theta)\pmb M_x\right)\right)^2\,,
\end{equation}
The quantum Fisher information (QFI), denoted as $\mathcal{F}_\theta$, is a measurement-independent quantity obtained from the maximum of all possible measurements, defined as \cite{2}

\begin{equation}
 F_\theta \equiv Tr (\pmb\rho_\theta\pmb L_{\rho_\theta}^2)=Tr( \pmb L_{\pmb\rho_\theta} \frac{\partial\pmb\rho_\theta}{\partial \theta})\,.
\end{equation}

After diagonalize the density matrix as $\pmb\rho_\theta=\sum_i \rho_i|i\rangle\langle i|$, the Fisher information can be reformulated as 
\begin{equation}
 F_\theta = \sum_i\frac{(\partial_\theta \rho_i)^2}{\rho_i} +2 \sum_{i\neq j}\frac{(\rho_i-\rho_j)^2}{\rho_i+\rho_j}|\langle \rho_i|\partial_\theta \rho_i\rangle|^2.
\end{equation}
where summations are taken over all $\rho_i\neq0$ and $\rho_i+\rho_j\neq0$.  For pure states, it simplifies to $F_\theta =4(\langle \partial_\theta\rho_i|\partial_\theta \rho_i\rangle-|\langle \rho_i|\partial_\theta \rho_i\rangle|^2)$. In the case of a single qubit system, the quantum Fisher information can be further simplified
\begin{equation}
F_\theta=\left\{
\begin{aligned}
|\partial_\theta \pmb r|^2 \quad\quad |\pmb r|=1 \\
|\partial_\theta \pmb r|^2+\frac{(\pmb r\cdot \partial_\theta \pmb r)^2}{1-|\pmb r|^2} \quad\quad |\pmb r|\neq 1
\end{aligned}
\right.
\end{equation}
where $\pmb r$ is the Bloch vector of the qubit.
Using the definition of the quantum Fisher information, an analogous inequality to the Cramér-Rao bound can be derived, known as the quantum Cramér-Rao bound. It is expressed as:
\begin{equation}
  \text{Var}(\theta_\text{est}) \geq \frac{1}{M \mathcal{F}_\theta}\,.
\end{equation}
where M is the number of independent identical POVM measurements on the systems with the same state. The measurements that achieve the quantum Cram\'er-Rao bound are termed as optimal distinguishing measurements \cite{1,2}.


\begin{figure}[!ht]
    \centering
\includegraphics[width=7 in]{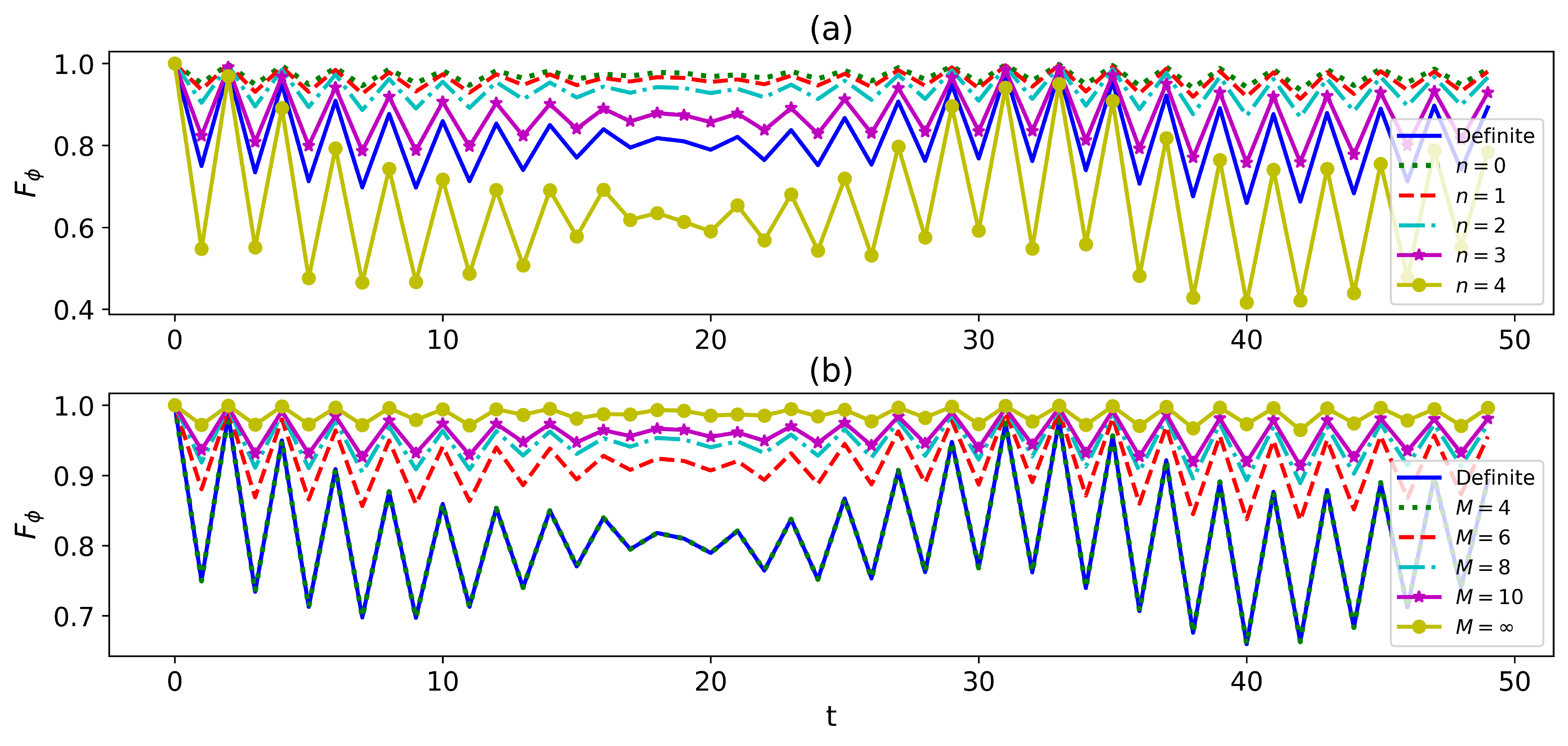}
\caption{\label{fig6} The variations of quantum Fisher information with the number of phase shifts when $M=10$ (a), with the number of paths when $n=1$ (b). The parameters in our analysis are $\omega_1=2, h=1.2, s=0.8$. The temperature is uniform and $T=0.3$. The single bath is composed of $N=100$ interacting qubits. }
\end{figure}
We examine the parameter estimation for a single qubit system. The parameters to be estimated can be encoded by a phase gate $\pmb U(\phi)=(|0\rangle\langle0|+e^{i\phi}|1\rangle\langle1|)$ acting on the system. And then let the system evolve in the indefinite environment introduced by the interferometer. To obtain the optimal Fisher information, it is beneficent to choose $\frac{1}{\sqrt{2}}(|0\rangle+|1\rangle)$ as an input state for the system qubit. In Fig. \ref{fig6}, we plot the dynamics of the quantum Fisher information for both definite and indefinite environments. The quantum Fisher information oscillates over time for both cases, but the indefinite environment can significantly improve the quantum Fisher information if the selective measurement is chosen properly. However, improperly chosen measurements can lead to even less quantum Fisher information, as shown by the yellow solid line with circle markers in Fig. \ref{fig6}(a). Notably, the effect of the indefiniteness is conspicuous for the case $n=0$. For the fixed finite phase shifts, the quantum Fisher information is significantly enhanced by increasing the paths as shown in Fig.\ref{fig6}(b). In particular, the Fisher information fluctuates very little and is very close to 1 if $M=\infty$.
Therefore, our analysis suggests that it is possible to enhance quantum parameter estimation by making the environment indefinite.

\section{Conclusion and Discussion}\label{Conclusion and Discussion}

In conclusion, we have analytically solved the central spin model to investigate the impact of indefinite environments on quantum systems. Our analysis reveals that quantum correlations can be boosted by introducing indefiniteness into the environments. However, if the path state is subject to decoherence, the interference between the different environments is lost quickly, and only a statistical mixture of the different environments remains. In this case, the dynamics of the system may not be significantly affected, especially if all the environments are uniform. We also considered the potential applications of our analysis, such as teleportation and parameter estimation. Our analysis shows that the indefiniteness of the environments can enhance both the fidelity and quantum Fisher information, indicating that it may improve the quantum characteristics in a wide range of applications. Notably, the central spin model can be realized using state-of-the-art experimental techniques, making it possible to test the indefiniteness of the qubit environments in the laboratory.

Finally, we discuss how to quantify the indefiniteness of the environments and why it impacts the evolution of the system. Without the loss of generality, we consider the scenario of control qubit. The system is a single qubit. One possible choice for quantifying indefiniteness is the coherence of the control qubit. However, this measure is not suitable since it excludes cases in which the coherence is absent but the mixedness of the environment still causes the indefiniteness.  Another possible candidate is the purity of the control qubit. However, this measure is also not appropriate since both the states $|0\rangle$ and $\frac{1}{\sqrt{2}}(|0\rangle+|1\rangle)$ share the maximum purity. Therefore, a more nuanced approach is required to define indefiniteness. Our starting point is the wave-particle-entanglement-ignorance (WPEI) relationship. The principle of complementarity, which is a defining characteristic of quantum objects, is exemplified by wave-particle duality \cite{WWK1979}. The wave-particle-entanglement (WPE) complementarity is an extension of the complementarity that combines wave-particle duality with quantum entanglement, initially developed for two-qubit systems \cite{JM10}. For mixed states of the two-qubit systems, the WPE complementarity was further extended to include another piece that characterizes ignorance, forming the WPEI complementarity \cite{TTE05, WW2020}.

\begin{figure}[!ht]
    \centering
\includegraphics[width=7 in]{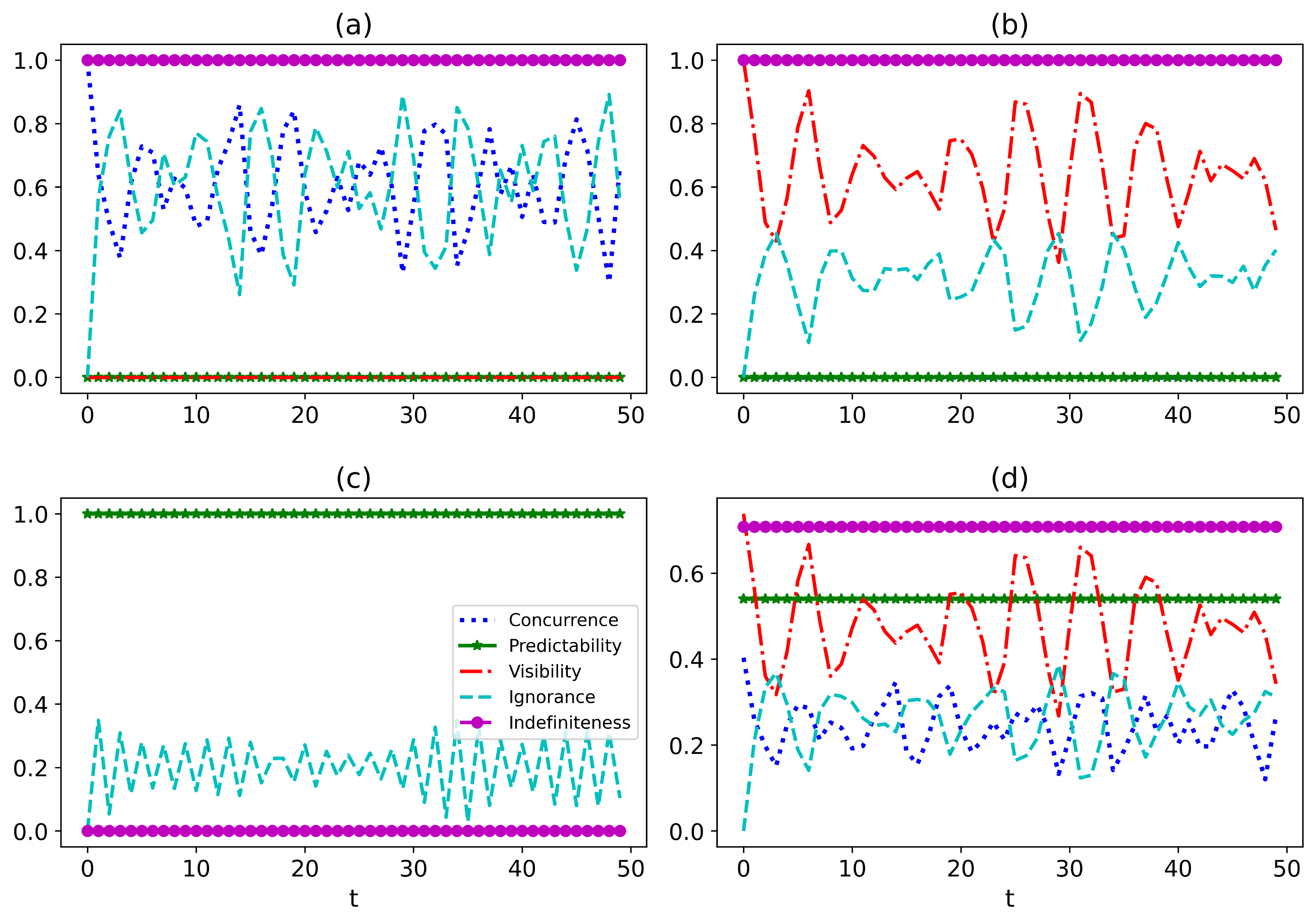}
\caption{\label{fig7} The variations of entanglement, predictability of the control qubit, visibility of the control qubit, ignorance, and indefiniteness w.r.t. time for the initial state $|\psi\rangle=\cos(\frac{\alpha}{2})|00\rangle+\cos(\frac{\theta}{2})\sin(\frac{\alpha}{2})|10\rangle+\sin(\frac{\theta}{2})\sin(\frac{\alpha}{2})|11\rangle$, where $\alpha$ and $\theta$ take on specific values for each of the four cases we consider: (a) $\alpha=\frac{\pi}{2},\theta=\pi$, (b) $\alpha=\frac{\pi}{2},\theta=0$, (c) $\alpha=\pi,\theta=\pi$, and (d) $\alpha=1,\theta=1$.  We analyze the variations of these parameters over time using a set of specific parameters, including $\omega_1=2$, $h=1.2$, and $s=0.8$. The temperatures are set to $T_0=0.1$ and $T_1=0.8$, and the single bath is composed of $N=100$ interacting qubits.}
\end{figure}

The WPEI complementarity provides a useful framework for quantifying the indefiniteness of the environment and understanding its impact on the evolution of the system. The complementarity is given by the expression $\frac{P_1^2+P_2^2+V_1^2+V_2^2}{2}+\mathscr{C}^2+\eta=1$ \cite{TTE05,WW2020}, where $P_i=|\rho_{i_{00}}-\rho_{i_{11}}|$ quantifies the path predictability for the i-th qubit, measuring the particleness of the system. The visibility of interference fringes measuring the wave property for the i-th qubit is given by $V_i=2|\rho_{i_{01}}|$. The concurrence $\mathscr{C}$ and the ignorance $\eta$ are also included in the expression, accounting for the entanglement and the lost information from the total system to the environment. We suggest that the indefiniteness of the environment can be quantified as $\mathscr{I}=1-P_1^2$. From the WPEI complementarity, we see that the indefiniteness is closely related to both local and non-local properties of the total system. To illustrate the impact of indefinitenesson the evolution of the system, we plot the entanglement, predictability, visibility, ignorance, and indefiniteness in Fig. \ref{fig7}. The plots show that the indefiniteness and the predictability are constant in our model once the initial state is determined, which is consistent with our setting. In Fig. \ref{fig7}(c), the indefiniteness is vanishing, representing the ordinary open quantum system dynamics. However, this is not the case for the other plots. Different levels of indefiniteness lead to distinct behavior of the local and non-local properties of the system. For example, the indefinite environment affects the system by affecting the entanglement between the system and the control qubit in Fig. \ref{fig7}(a), and by affecting the visibility of the control qubit in Fig. \ref{fig7}(b). In Fig. \ref{fig7}(d), the indefiniteness affects both the local and non-local properties of the system. The indefiniteness suggests a bound for the evolution of both local and non-local properties. This explains why the indefiniteness of the environment impacts the dynamics of the system. By developing a better understanding of the behavior of quantum systems in indefinite environments, we provide insights into the design of more efficient and accurate quantum technologies.

\appendix
\section{The evolution of the two-qubit system in indefinite environments}\label{appendix_A}

In this section, we give the details of the evolution of the two-qubit system in indefinite environments. Following the procedures described in Sec.\ref{Model and dynamics} in the main text, after lengthy but straightforwards calculation we derive,

Case A: $i=i',j=j'$
\begin{equation}\begin{split}
\rho_{11}(t)&=\frac{1}{Z_{ij}}\sum_{n_i}^N \sum_{n_j}^N exp(-\beta_js(n_j(1-\frac{ n_j-1}{N})-\frac{1}{2}))exp(-\beta_i s( n_i(1-\frac{n_i-1}{N})-\frac{1}{2}))\\
&(|A_1(n_i,n_j,t)|^2\rho_{11}(0) + n_j|K_1(n_i,n_j,t)|^2\rho_{22}(0) + n_i|I_1(n_i,n_j,t)|^2\rho_{33}(0)),\\
\rho_{22}(t)&=\frac{1}{Z_{ij}}\sum_{n_i}^N \sum_{n_j}^N exp(-\beta_j s( n_j(1-\frac{ n_j-1}{N})-\frac{1}{2}))exp(-\beta_i s( n_i(1-\frac{ n_i-1}{N})-\frac{1}{2})) \\
&(|J_1(n_i,n_j,t)|^2 \rho_{22}(0)+(n_j+1)|B_1(n_i,n_j,t)|^2\rho_{11}(0)+n_i|F_1(n_i,n_j,t)|^2\rho_{44}(0)),\\
\rho_{33}(t)&=\frac{1}{Z_{ij}}\sum_{n_i}^N \sum_{n_j}^N  exp(-\beta_j s( n_j(1-\frac{ n_j-1}{N})-\frac{1}{2}))exp(-\beta_i s( n_i(1-\frac{ n_i-1}{N})-\frac{1}{2})) \\
&(|G_1(n_i,n_j,t)|^2 \rho_{33}(0) +(n_i+1)|C_1(n_i,n_j,t)|^2\rho_{11}(0)+ n_j|E_1(n_i,n_j,t)|^2\rho_{44}(0)),\\
\rho_{44}(t)&=\frac{1}{Z_{ij}}\sum_{n_i}^N \sum_{n_j}^N  exp(-\beta_j s( n_j(1-\frac{ n_j-1}{N})-\frac{1}{2}))exp(-\beta_i s( n_i(1-\frac{ n_i-1}{N})-\frac{1}{2}))\\
&(|D_1(n_i,n_j,t)|^2 \rho_{44}(0) +(n_i+1)|L_1(n_i,n_j,t)|^2\rho_{22}(0) + (n_j+1)|H_1(n_i,n_j,t)|^2\rho_{33}(0)),\\
\rho_{12}(t)&=\frac{1}{Z_{ij}}\sum_{n_i}^N \sum_{n_j}^N exp(-\beta_j s( n_j(1-\frac{ n_j-1}{N})-\frac{1}{2}))exp(-\beta_i s( n_i(1-\frac{ n_i-1}{N})-\frac{1}{2})) \\
&(A_1(n_i,n_j,t)J_1^*(n_i,n_j,t)\rho_{12}(0) + n_i I_1(n_i,n_j,t)F_1^*(n_i,n_j,t)\rho_{34}(0)),\\
\rho_{13}(t)&=\frac{1}{Z_{ij}}\sum_{n_i}^N \sum_{n_j}^N  exp(-\beta_j s( n_j(1-\frac{ n_j-1}{N})-\frac{1}{2}))exp(-\beta_i s( n_i(1-\frac{ n_i-1}{N})-\frac{1}{2})) \\
&(A_1(n_i,n_j,t)G_1^*(n_i,n_j,t)\rho_{13}(0) + n_j K_1(n_i,n_j,t)E_1^*(n_i,n_j,t)\rho_{24}(0)),\\
\rho_{14}(t)&=\frac{1}{Z_{ij}}\sum_{n_i}^N \sum_{n_j}^N  exp(-\beta_j s( n_j(1-\frac{ n_j-1}{N})-\frac{1}{2}))exp(-\beta_i s( n_i(1-\frac{ n_i-1}{N})-\frac{1}{2})) ((A_1(n_i,n_j,t)D_1^*(n_i,n_j,t)\rho_{14}(0)),\\
\rho_{23}(t)&=\frac{1}{Z_{ij}}\sum_{n_i}^N \sum_{n_j}^N  exp(-\beta_j s( n_j(1-\frac{ n_j-1}{N})-\frac{1}{2}))exp(-\beta_i s( n_i(1-\frac{ n_i-1}{N})-\frac{1}{2})) (J_1(n_i,n_j,t)G_1^*(n_i,n_j,t)\rho_{23}(0) ),\\
\rho_{24}(t)&=\frac{1}{Z_{ij}}\sum_{n_i}^N \sum_{n_j}^N  exp(-\beta_j s( n_j(1-\frac{ n_j-1}{N})-\frac{1}{2}))exp(-\beta_i s( n_i(1-\frac{ n_i-1}{N})-\frac{1}{2})) \\
&(J_1(n_i,n_j,t)D_1^*(n_i,n_j,t)\rho_{24}(0) +n_j B_1(n_i,n_j,t)H_1^*(n_i,n_j,t)\rho_{13}(0)),\\
\rho_{34}(t)&=\frac{1}{Z_{ij}}\sum_{n_i}^N \sum_{n_j}^N  exp(-\beta_j s( n_j(1-\frac{ n_j-1}{N})-\frac{1}{2}))exp(-\beta_i s( n_i(1-\frac{ n_i-1}{N})-\frac{1}{2}))\\ 
&(G_1(n_i,n_j,t)D_1^*(n_i,n_j,t)\rho_{34}(0)+(n_i+1) C_1(n_i,n_j,t)L_1^*(n_i,n_j,t) \rho_{12}(0)).
\end{split}
\end{equation}
where $Z_{ij}=\sum_{n_i}^N \sum_{n_j}^N exp(-\beta_js(n_j(1-\frac{ n_j-1}{N})-\frac{1}{2}))exp(-\beta_i s( n_i(1-\frac{n_i-1}{N})-\frac{1}{2}))$.

Case B: $i= i', j\neq j'$ or $i\neq i', j= j'$. We only show the result of case $i= i', j\neq j'$. Another case is similar.

\begin{equation}\begin{split}
\rho_{11}(t)&=\frac{1}{Z_{ijj'}}\sum_{n_{i},n_j,n_{j'}}exp(-\beta_{j'}s(n_{j'}(1-\frac{ n_{j'}-1}{N})-\frac{1}{2}))exp(-\beta_js(n_j(1-\frac{ n_j-1}{N})-\frac{1}{2}))exp(-\beta_i s( n_i(1-\frac{n_i-1}{N})-\frac{1}{2}))\\
&(A_1(n_i,n_j,t)A_1(n_i,n_{j'},t)^*\rho_{11}(0) + n_iI_1(n_i,n_j,t)I_1(n_i,n_{j'},t)\rho_{33}(0)),\\
\rho_{22}(t)&=\frac{1}{Z_{ijj'}}\sum_{n_{i},n_j,n_{j'}}exp(-\beta_{j'}s(n_{j'}(1-\frac{ n_{j'}-1}{N})-\frac{1}{2}))exp(-\beta_js(n_j(1-\frac{ n_j-1}{N})-\frac{1}{2}))exp(-\beta_i s( n_i(1-\frac{n_i-1}{N})-\frac{1}{2})) \\
&(J_1(n_i,n_j,t)J_1(n_i,n_{j'},t)^* \rho_{22}(0)+n_iF_1(n_i,n_j,t)F_1(n_i,n_{j'},t)\rho_{44}(0)),\\
\rho_{33}(t)&=\frac{1}{Z_{ijj'}}\sum_{n_{i},n_j,n_{j'}}exp(-\beta_{j'}s(n_{j'}(1-\frac{ n_{j'}-1}{N})-\frac{1}{2}))exp(-\beta_js(n_j(1-\frac{ n_j-1}{N})-\frac{1}{2}))exp(-\beta_i s( n_i(1-\frac{n_i-1}{N})-\frac{1}{2}))\\
&(G_1(n_i,n_j,t)G_1(n_i,n_{j'},t)\rho_{33}(0) +(n_i+1)C_1(n_i,n_j,t)C_1(n_i,n_{j'},t)\rho_{11}(0)),\\
\rho_{44}(t)&=\frac{1}{Z_{ijj'}}\sum_{n_{i},n_j,n_{j'}}exp(-\beta_{j'}s(n_{j'}(1-\frac{ n_{j'}-1}{N})-\frac{1}{2}))exp(-\beta_js(n_j(1-\frac{ n_j-1}{N})-\frac{1}{2}))exp(-\beta_i s( n_i(1-\frac{n_i-1}{N})-\frac{1}{2}))\\
&(D_1(n_i,n_j,t)D_1(n_i,n_{j'},t) \rho_{44}(0) +(n_i+1)L_1(n_i,n_j,t)L_1(n_i,n_{j'},t)\rho_{22}(0)),\\
\rho_{12}(t)&=\frac{1}{Z_{ijj'}}\sum_{n_{i},n_j,n_{j'}}exp(-\beta_{j'}s(n_{j'}(1-\frac{ n_{j'}-1}{N})-\frac{1}{2}))exp(-\beta_js(n_j(1-\frac{ n_j-1}{N})-\frac{1}{2}))exp(-\beta_i s( n_i(1-\frac{n_i-1}{N})-\frac{1}{2})) \\
&(A_1(n_i,n_j,t)J_1^*(n_i,n_{j'}t)\rho_{12}(0) + n_i I_1(n_i,n_j,t)F_1^*(n_i,n_{j'},t)\rho_{34}(0)),\\
\rho_{13}(t)&=\frac{1}{Z_{ijj'}}\sum_{n_{i},n_j,n_{j'}}exp(-\beta_{j'}s(n_{j'}(1-\frac{ n_{j'}-1}{N})-\frac{1}{2}))exp(-\beta_js(n_j(1-\frac{ n_j-1}{N})-\frac{1}{2}))exp(-\beta_i s( n_i(1-\frac{n_i-1}{N})-\frac{1}{2})) \\
&(A_1(n_i,n_j,t)G_1^*(n_i,n_{j'},t)\rho_{13}(0)),\\
\rho_{14}(t)&=\frac{1}{Z_{ijj'}}\sum_{n_i}^N \sum_{n_j}^N  exp(-\beta_{j'}s(n_{j'}(1-\frac{ n_{j'}-1}{N})-\frac{1}{2}))exp(-\beta_js(n_j(1-\frac{ n_j-1}{N})-\frac{1}{2}))exp(-\beta_i s( n_i(1-\frac{n_i-1}{N})-\frac{1}{2}))  \\
&(A_1(n_i,n_j,t)D_1^*(n_i,n_{j'},t)\rho_{14}(0)),\\
\rho_{23}(t)&=\frac{1}{Z_{ijj'}}\sum_{n_{i},n_j,n_{j'}}exp(-\beta_{j'}s(n_{j'}(1-\frac{ n_{j'}-1}{N})-\frac{1}{2}))exp(-\beta_js(n_j(1-\frac{ n_j-1}{N})-\frac{1}{2}))exp(-\beta_i s( n_i(1-\frac{n_i-1}{N})-\frac{1}{2}))\\
&(J_1(n_i,n_j,t)G_1^*(n_i,n_{j'},t)\rho_{23}(0)),\\
\rho_{24}(t)&=\frac{1}{Z_{ijj'}}\sum_{n_{i},n_j,n_{j'}}exp(-\beta_{j'}s(n_{j'}(1-\frac{ n_{j'}-1}{N})-\frac{1}{2}))exp(-\beta_js(n_j(1-\frac{ n_j-1}{N})-\frac{1}{2}))exp(-\beta_i s( n_i(1-\frac{n_i-1}{N})-\frac{1}{2})) \\
&(J_1(n_i,n_j,t)D_1^*(n_i,n_{j'},t)\rho_{24}(0)),\\
\rho_{34}(t)&=\frac{1}{Z_{ijj'}}\sum_{n_{i},n_j,n_{j'}}exp(-\beta_{j'}s(n_{j'}(1-\frac{ n_{j'}-1}{N})-\frac{1}{2}))exp(-\beta_js(n_j(1-\frac{ n_j-1}{N})-\frac{1}{2}))exp(-\beta_i s( n_i(1-\frac{n_i-1}{N})-\frac{1}{2}))\\ 
&(G_1(n_i,n_j,t)D_1^*(n_i,n_{j'},t)\rho_{34}(0)+(n_i+1) C_1(n_i,n_j,t)L_1^*(n_i,n_{j'},t) \rho_{12}(0)).
\end{split}
\end{equation}

where $Z_{ijj'}=\sum_{n_{i},n_j,n_{j'}}exp(-\beta_{j'}s(n_{j'}(1-\frac{ n_{j'}-1}{N})-\frac{1}{2}))exp(-\beta_js(n_j(1-\frac{ n_j-1}{N})-\frac{1}{2}))exp(-\beta_i s( n_i(1-\frac{n_i-1}{N})-\frac{1}{2}))$.

Case C:$i\neq i', j\neq j'$.

\begin{equation}\begin{split}
\rho_{11}(t)&=\frac{1}{Z_{iji'j'}}\sum_{n_{i},n_j,n_{i'},n_{j'}}exp(-\beta_i s( n_i(1-\frac{n_i-1}{N})-\frac{1}{2}))exp(-\beta_js(n_j(1-\frac{ n_j-1}{N})-\frac{1}{2}))\\
&exp(-\beta_{i'}s(n_{i'}(1-\frac{ n_{i'}-1}{N})-\frac{1}{2}))exp(-\beta_{j'}s(n_{j'}(1-\frac{ n_{j'}-1}{N})-\frac{1}{2}))A_1(n_i,n_j,t)A_1(n_{i'},n_{j'},t)^*\rho_{11}(0),\\
\rho_{22}(t)&=\frac{1}{Z_{iji'j'}}\sum_{n_{i},n_j,n_{i'},n_{j'}}exp(-\beta_i s( n_i(1-\frac{n_i-1}{N})-\frac{1}{2}))exp(-\beta_js(n_j(1-\frac{ n_j-1}{N})-\frac{1}{2}))\\
&exp(-\beta_{i'}s(n_{i'}(1-\frac{ n_{i'}-1}{N})-\frac{1}{2}))exp(-\beta_{j'}s(n_{j'}(1-\frac{ n_{j'}-1}{N})-\frac{1}{2}))J_1(n_i,n_j,t)J_1(n_{i'},n_{j'},t)^* \rho_{22}(0)\\
\rho_{33}(t)&=\frac{1}{Z_{iji'j'}}\sum_{n_{i},n_j,n_{i'},n_{j'}}exp(-\beta_i s( n_i(1-\frac{n_i-1}{N})-\frac{1}{2}))exp(-\beta_js(n_j(1-\frac{ n_j-1}{N})-\frac{1}{2}))\\
&exp(-\beta_{i'}s(n_{i'}(1-\frac{ n_{i'}-1}{N})-\frac{1}{2}))exp(-\beta_{j'}s(n_{j'}(1-\frac{ n_{j'}-1}{N})-\frac{1}{2}))G_1(n_i,n_j,t)G_1(n_{i'},n_{j'},t)\rho_{33}(0),\\
\rho_{44}(t)&=\frac{1}{Z_{iji'j'}}\sum_{n_{i},n_j,n_{i'},n_{j'}}exp(-\beta_i s( n_i(1-\frac{n_i-1}{N})-\frac{1}{2}))exp(-\beta_js(n_j(1-\frac{ n_j-1}{N})-\frac{1}{2}))\\
&exp(-\beta_{i'}s(n_{i'}(1-\frac{ n_{i'}-1}{N})-\frac{1}{2}))exp(-\beta_{j'}s(n_{j'}(1-\frac{ n_{j'}-1}{N})-\frac{1}{2}))D_1(n_i,n_j,t)D_1(n_{i'},n_{j'},t) \rho_{44}(0) ,\\
\rho_{12}(t)&=\frac{1}{Z_{iji'j'}}\sum_{n_{i},n_j,n_{i'},n_{j'}}exp(-\beta_i s( n_i(1-\frac{n_i-1}{N})-\frac{1}{2}))exp(-\beta_js(n_j(1-\frac{ n_j-1}{N})-\frac{1}{2}))\\
&exp(-\beta_{i'}s(n_{i'}(1-\frac{ n_{i'}-1}{N})-\frac{1}{2}))exp(-\beta_{j'}s(n_{j'}(1-\frac{ n_{j'}-1}{N})-\frac{1}{2}))A_1(n_i,n_j,t)J_1^*(n_{i'},n_{j'}t)\rho_{12}(0),\\
\rho_{13}(t)&=\frac{1}{Z_{iji'j'}}\sum_{n_{i},n_j,n_{i'},n_{j'}}exp(-\beta_i s( n_i(1-\frac{n_i-1}{N})-\frac{1}{2}))exp(-\beta_js(n_j(1-\frac{ n_j-1}{N})-\frac{1}{2}))\\
&exp(-\beta_{i'}s(n_{i'}(1-\frac{ n_{i'}-1}{N})-\frac{1}{2}))exp(-\beta_{j'}s(n_{j'}(1-\frac{ n_{j'}-1}{N})-\frac{1}{2}))A_1(n_i,n_j,t)G_1^*(n_{i'},n_{j'},t)\rho_{13}(0),\\
\rho_{14}(t)&=\frac{1}{Z_{iji'j'}}\sum_{n_{i},n_j,n_{i'},n_{j'}}exp(-\beta_i s( n_i(1-\frac{n_i-1}{N})-\frac{1}{2}))exp(-\beta_js(n_j(1-\frac{ n_j-1}{N})-\frac{1}{2}))\\
&exp(-\beta_{i'}s(n_{i'}(1-\frac{ n_{i'}-1}{N})-\frac{1}{2}))exp(-\beta_{j'}s(n_{j'}(1-\frac{ n_{j'}-1}{N})-\frac{1}{2}))A_1(n_i,n_j,t)D_1^*(n_{i'},n_{j'},t)\rho_{14}(0),\\
\rho_{23}(t)&=\frac{1}{Z_{iji'j'}}\sum_{n_{i},n_j,n_{i'},n_{j'}}exp(-\beta_i s( n_i(1-\frac{n_i-1}{N})-\frac{1}{2}))exp(-\beta_js(n_j(1-\frac{ n_j-1}{N})-\frac{1}{2}))\\
&exp(-\beta_{i'}s(n_{i'}(1-\frac{ n_{i'}-1}{N})-\frac{1}{2}))exp(-\beta_{j'}s(n_{j'}(1-\frac{ n_{j'}-1}{N})-\frac{1}{2}))J_1(n_i,n_j,t)G_1^*(n_{i'},n_{j'},t)\rho_{23}(0),\\
\rho_{24}(t)&=\frac{1}{Z_{iji'j'}}\sum_{n_{i},n_j,n_{i'},n_{j'}}exp(-\beta_i s( n_i(1-\frac{n_i-1}{N})-\frac{1}{2}))exp(-\beta_js(n_j(1-\frac{ n_j-1}{N})-\frac{1}{2}))\\
&exp(-\beta_{i'}s(n_{i'}(1-\frac{ n_{i'}-1}{N})-\frac{1}{2}))exp(-\beta_{j'}s(n_{j'}(1-\frac{ n_{j'}-1}{N})-\frac{1}{2}))J_1(n_i,n_j,t)D_1^*(n_{i'},n_{j'},t)\rho_{24}(0),\\
\rho_{34}(t)&=\frac{1}{Z_{iji'j'}}\sum_{n_{i},n_j,n_{i'},n_{j'}}exp(-\beta_i s( n_i(1-\frac{n_i-1}{N})-\frac{1}{2}))exp(-\beta_js(n_j(1-\frac{ n_j-1}{N})-\frac{1}{2}))\\
&exp(-\beta_{i'}s(n_{i'}(1-\frac{ n_{i'}-1}{N})-\frac{1}{2}))exp(-\beta_{j'}s(n_{j'}(1-\frac{ n_{j'}-1}{N})-\frac{1}{2}))G_1(n_i,n_j,t)D_1^*(n_{i'},n_{j'},t)\rho_{34}(0).
\end{split}
\end{equation}

where $Z_{iji'j'}=\sum_{n_{i},n_j,n_{i'},n_{j'}}exp(-\beta_{j'}s(n_{j'}(1-\frac{ n_{j'}-1}{N})-\frac{1}{2}))exp(-\beta_js(n_j(1-\frac{ n_j-1}{N})-\frac{1}{2}))exp(-\beta_{i'}s(n_{i'}(1-\frac{ n_{i'}-1}{N})-\frac{1}{2}))exp(-\beta_i s( n_i(1-\frac{n_i-1}{N})-\frac{1}{2}))$.

\section{The evolution of the system density matrix for the teleportation process}\label{appendix_B}

Here we present the full details of the density matrix during the teleportation.
\begin{equation}\begin{split}
\pmb \rho_{ABC}&=\frac{1}{2}\frac{1}{Z_0Z_1} \sum_{n_0}^N \sum_{n_1}^N exp(-\beta_0 s( n_0(1-\frac{ n_0-1}{N})-\frac{1}{2})))exp(-\beta_1 s( n_1(1-\frac{ n_1-1}{N})-\frac{1}{2})))\\
&(|c_0|^2|0\rangle\langle0|\otimes((|A_1(n_0,t)|^2|A_1(n_1,t)|^2+n_0n_1|D_1(n_0,t)|^2|D_1(n_1,t)|^2)|1\rangle\langle1|\otimes|1\rangle\langle1|\\
&+((n_1+1)|A_1(n_0,t)|^2|B_1(n_1,t)|^2 +(n_0)|D_1(n_0,t)|^2|C_1(n_1,t)|^2)|1\rangle\langle1|\otimes|0\rangle\langle0|\\
&+((n_0+1)|B_1(n_0,t)|^2 |A_1(n_1,t)|^2+(n_1)|C_1(n_0,t)|^2 |D_1(n_1,t)|^2)|0\rangle\langle 0|\otimes|1\rangle\langle1|\\
&+((n_0+1)(n_1+1)|B_1(n_0,t)|^2|B_1(n_1,t)|^2+|C_1(n_0,t)|^2|C_1(n_1,t)|^2)|0\rangle\langle 0|\otimes |0\rangle\langle 0|\\
&+A_1(n_0,t)*C_1^{*}(n_0,t)  A_1(n_1,t)*C_1^{*}(n_1,t) |1\rangle\langle0|\otimes|1\rangle\langle0|\\
&+C_1(n_0,t)*A_1^{*}(n_0,t)  C_1(n_1,t)*A_1^{*}(n_1,t) |0\rangle\langle1|\otimes|0\rangle\langle1|)\\
&+|c_1|^2|1\rangle\langle1|\otimes(|A_1(n_1,t)|^2|A_1(n_0,t)|^2+n_0n_1|D_1(n_1,t)|^2|D_1(n_0,t)|^2)|1\rangle\langle1|\otimes|1\rangle\langle1|\\
&+((n_0+1)|A_1(n_1,t)|^2|B_1(n_0,t)|^2 +(n_1)|D_1(n_1,t)|^2|C_1(n_0,t)|^2)|1\rangle\langle1|\otimes|0\rangle\langle0|\\
&+((n_1+1)|B_1(n_1,t)|^2 |A_1(n_0,t)|^2+(n_0)|C_1(n_1,t)|^2 |D_1(n_0,t)|^2)|0\rangle\langle 0|\otimes|1\rangle\langle1|\\
&+((n_1+1)(n_0+1)|B_1(n_1,t)|^2|B_1(n_0,t)|^2+|C_1(n_1,t)|^2|C_1(n_0,t)|^2)|0\rangle\langle 0|\otimes |0\rangle\langle 0|\\
&+(A_1(n_1,t)*C_1^{*}(n_1,t)  A_1(n_0,t)*C_1^{*}(n_0,t) |1\rangle\langle0|\otimes|1\rangle\langle0|)\\
&+ (C_1(n_1,t)*A_1^{*}(n_1,t) C_1(n_0,t)*A_1^{*}(n_0,t) |0\rangle\langle1|\otimes |0\rangle\langle1|)\\
&+c_1c_0^*|1\rangle\langle0|\otimes((|A_1(n_0,t)|^2|A_1(n_1,t)|^2+(n_0)(n_1) |D_1(n_0,t)|^2 |D_1(n_1,t)|^2)|1,1\rangle\langle1,1| \\
&+((n_0+1) |B_1(n_0,t)|^2|A_1(n_1,t)|^2+(n_1) |C_1(n_0,t)|^2|D_1(n_1,t)|^2)|1,0\rangle\langle0,1|\\
&+((n_1+1)|A_1(n_0,t)|^2|B_1(n_1,t)|^2+(n_0)|C_1(n_1,t)|^2|D_1(n_0,t))|^2|0,1\rangle\langle 1,0|\\
&+(|C_1(n_0,t)|^2|C_1(n_1,t)|^2+(n_0+1)(n_1+1) |B_1(n_0,t)|^2 |B_1(n_1,t)|^2)|0,0\rangle\langle0,0|\\
&+A_1(n_0,t)A_1(n_1,t)C_1^*(n_0,t)C_1^*(n_1,t) |1,1\rangle\langle0,0|\\
&+A_1^*(n_0,t)A_1^*(n_1,t)C_1(n_0,t)C_1(n_1,t) |0,0\rangle\langle1,1|)\\
&+c_0c_1^*|0\rangle\langle1|\otimes((|A_1(n_0,t)|^2|A_1(n_1,t)|^2+(n_0)(n_1) |D_1(n_0,t)|^2 |D_1(n_1,t)|^2)|1,1\rangle\langle1,1| \\
&+((n_1+1) |B_1(n_1,t)|^2|A_1(n_0,t)+(n_0)|C_1(n_1,t)|^2|D_1(n_0,t)|^2|^2)|1,0\rangle\langle0,1|\\
&+((n_0+1)|A_1(n_1,t)|^2|B_1(n_0,t)|^2+(n_1)|C_1(n_0,t)|^2|D_1(n_1,t))|^2|0,1\rangle\langle 1,0|\\
&+(|C_1(n_0,t)|^2|C_1(n_1,t)|^2+(n_0+1)(n_1+1) |B_1(n_0,t)|^2 |B_1(n_1,t)|^2)|0,0\rangle\langle0,0|\\
&+A_1(n_0,t)A_1(n_1,t)C_1^*(n_0,t)C_1^*(n_1,t)|1,1\rangle\langle0,0|\\
&+A_1^*(n_0,t)A_1^*(n_1,t)C_1(n_0,t)C_1(n_1,t)|0,0\rangle\langle1,1|))
\end{split}
\end{equation}

where $Z_0=\sum_{n_0}^N exp(-\beta_0 s( n_0(1-\frac{ n_0-1}{N})-\frac{1}{2})))$ and $Z_1=\sum_{n_1}^N exp(-\beta_1 s( n_1(1-\frac{ n_1-1}{N})-\frac{1}{2})))$.

\section{The maximum fidelity of teleporting certain states}

We calculate the maximum fidelity of teleporting certain states using the general mixed states as a resource. The post-measurement states are $\{\pmb\rho_{B}^{0}, \pmb\rho_{B}^{1}, \pmb\rho_{B}^{2}, \pmb\rho_{B}^{3}\}$ with probability $\{P^{0}, P^{1}, P^{2}, P^{3}\}$. It is helpful to rewrite them in
Bloch representation,
\begin{equation}
\pmb\rho_{B}^{k}   = \frac{1}{2}(\pmb I + \pmb n^{k}\cdot\pmb\sigma )
\end{equation}

The problem that finds the unitary transformation $\pmb U_{k}$ such that $F=Tr(\pmb U_{k}\pmb\rho_{B}^{k}\pmb U^{k\dagger}\pmb \rho_{C} )$ maximize is equivalent to find the rotation $\pmb O^k$ such that

\begin{equation}
F_k=  \frac{1}{2}(1 + \pmb O^k\pmb n^{k}\cdot \pmb n^C),
\end{equation}
maximize $F_k$. The rotation matrix can be given by Rodrigues rotation formula

\begin{equation}
\pmb O^k= \pmb I + \sin(\theta) \pmb K + (1 - \cos(\theta)) \pmb K^2
\end{equation}
where $\theta=\arccos(\pmb n^{k}\cdot \pmb n^C / (||\pmb n^{k}|| ||\pmb n^C||))$, $\pmb K=\pmb n^{k}\times \pmb n^C$. Finally, the total fidelity is 

\begin{equation}
F= \sum_{k}P^kF_k.
\end{equation}

\end{document}